\documentclass[twocolumn]{aastex631}

\usepackage{graphicx}% Include figure files
\usepackage{amsmath}
\usepackage{rviewport}
\usepackage{xcolor}
\usepackage{dcolumn}% Align table columns on decimal point
\usepackage{bm}% bold math
\usepackage{hyperref}% add hypertext capabilities
\usepackage{xspace}
\usepackage{mathtools}
\usepackage{tikz}
\usepackage[caption=false]{subfig}
\usetikzlibrary{shapes}

\makeatletter
\newcommand\mathcircled[1]{%
  \mathpalette\@mathcircled{#1}%
}
\newcommand\@mathcircled[2]{%
  \tikz[baseline=(math.base)] \node[draw,ellipse,inner sep=1pt] (math) {$\m@th#1#2$};%
}
\makeatother

\begin{document}

\title{The large-scale anisotropy and flux (de)magnification\\ of ultra-high-energy cosmic rays in the Galactic magnetic field}

\author[0000-0003-4005-0857]{Teresa Bister}
\email{teresa.bister@ru.nl}
\affiliation{Institute for Mathematics, Astrophysics and Particle Physics, Radboud University Nijmegen, Nijmegen, The Netherlands}
\affiliation{Nationaal Instituut voor Subatomaire Fysica (NIKHEF), Science Park, Amsterdam, The Netherlands}

\author[0000-0003-2417-5975]{Glennys R. Farrar}
\email{gf25@nyu.edu}
\affiliation{Center for Cosmology and Particle Physics, New York University, New York, NY 10003, USA}%

\author[0000-0002-7651-0272]{Michael Unger}
\email{michael.unger@kit.edu}
\affiliation{Institute for Astroparticle Physics, Karlsruhe Institute of Technology (KIT), Karlsruhe, Germany}
\affiliation{Institutt for fysikk, Norwegian University of Science and Technology (NTNU), Trondheim, Norway}%

\begin{abstract}
We calculate the arrival direction distribution of ultra-high-energy cosmic rays (UHECRs) with a new suite of models of the Galactic magnetic field (GMF), assuming sources follow the large-scale structure of the Universe. Compared to previous GMF models, the amplitude of the dipole component of the UHECR arrival flux is significantly reduced. We find that the reduction is due to the accidentally coinciding position of the peak of the extragalactic UHECR flux and the boundary of strong flux demagnification due to the GMF toward the central region of the Galaxy.
This serendipitous sensitivity of UHECR anisotropies to the GMF model will be a powerful probe of the source distribution as well as Galactic and extragalactic magnetic fields.  Demagnification by the GMF also impacts visibility of some popular source candidates.
\end{abstract}


\section{Introduction} \label{sec:intro}
\textit{Ultra-high-energy cosmic rays} (UHECRs) are the highest energetic particles measured at Earth, with energies from $10^{18}\,\mathrm{eV}$ to beyond $10^{20}\,\mathrm{eV}$. Their origin remains unclear, mainly because UHECRs are charged nuclei which are deflected by cosmic magnetic fields during their propagation from the sources to Earth. Hence, the directions of sources can only be reconstructed from the UHECR arrival directions when magnetic field deflections are appropriately accounted for. A large part of the effect comes from the \textit{Galactic magnetic field} (GMF) of the Milky Way, which has a field strength of order $\mu\mathrm{G}$ extending over tens of kpc.
The mass composition of UHECRs becomes heavier with increasing energy $E$~\citep{PierreAuger:2014gko}, with a relatively narrow range of rigidities $\mathcal{R} \equiv E/Ze \approx 5\,\mathrm{EV}$~\citep{BF23}, for energies $E\gtrsim8\,\mathrm{EeV}$ relevant for this work. The Larmor radius of UHECRs is $\sim5.5$ kpc $(\mathcal{R}/5\,$EV)/(B/$\mu$G) - hence the GMF has a sizable impact on the propagation of UHECRs.

In recent years, the Jansson-Farrar GMF model from 2012 (\texttt{JF12})~\citep{jansson_galactic_2012, jansson_new_2012} has been used to test hypotheses about the sources of UHECRs from irregularities in the UHECR arrival directions (e.g.~\citet{ding_imprint_2021, BF23, Eichmann:2022ias, Globus_2019, Globus:2022qcr, eichmann_2020, allard_what_2022}),
but the robustness of the conclusions of these studies was difficult to assess due to the absence of realistic alternative GMF models which also fit the full data. Several Galactic magnetic field models fit only Faraday rotation measures (RMs)  but not polarized synchrotron emission, and some models only fit for the disk field, see~\citet{Jaffe:2019iuk} and references therein; see also~\citet{Korochkin:2024yit} for a recent study of the GMF halo component using high-latitude RMs and polarized synchrotron emission. For comparisons of UHECR anisotropy predictions by some of those models and \texttt{JF12} see, e.g.~\citet{di_Matteo_2018, Erdmann_2016, allard_what_2022}.

Many of the references given above aim at modeling the \textit{dipole} - the only currently significant anisotropy in the arrival directions of UHECRs at $E>8\,\mathrm{EeV}$~\citep{Auger_dipole_2017_Science, Auger_dipole_2018}. It has a magnitude of $\sim7.3\%$ and a current significance of $6.9\sigma$~\citep{Golup_ADs_2023} in the field of view of the Pierre Auger Observatory~\citep{auger_2015}. All higher multipole moments, however, are compatible with isotropy, according to the joint analysis of the Pierre Auger and Telescope Array (TA) collaborations~\citep{Auger_dipole_2018, Caccianiga_ICRC2023}.
Using the \texttt{JF12} model for the Galactic magnetic field, it was verified by~\citep{BF23} that the dipole amplitude including its energy dependence can be explained rather well if UHECR sources follow the extragalactic matter distribution and hence the large-scale structure (LSS) of the Universe, while the measured dipole direction is only roughly right.  Additionally, constraints on the source number density and the extragalactic magnetic field smearing were derived by requiring that all higher multipoles are compatible with isotropy.

Recently, new modeling of the GMF has become available in~\citet{UF23}, hereafter \texttt{UF23}. In addition to being based on the latest astronomical data, \texttt{UF23} provides a suite of models using a variety of improved functional forms for the field and for the thermal and cosmic-ray electron densities which are needed to predict the observables (RMs and polarized synchrotron emission), intended to encapsulate the uncertainty in the coherent GMF. In this work, we discuss the predictions of the large-scale distribution of arriving UHECRs according to the new \texttt{UF23} GMF models.  We show in particular the important influence of the magnification and demagnification effect of the Galactic magnetic field.

The relevance of the anisotropic (de)magnification due to the GMF is amplified by the fact that due to energy losses in propagation, the ``UHECR illumination" of the Galaxy is quite inhomogeneous. If the source density is high enough that the source distribution reflects the distribution of matter (and UHECRs are not magnetically trapped within nearby Galaxy clusters~\citep{Condorelli_2023}), the flux of UHECRs above 8 EeV arriving at the Galaxy will be considerably enhanced in the direction of the Virgo cluster and Great Attractor (see Fig.~2 of~\citet{BF23}). As we show, the alignment of the quite concentrated illumination map with the boundary of GMF demagnification has a strong impact on the predicted dipole magnitude and direction, enabling greater sensitivity in probing the various contributing factors.

\section{Dependence of the anisotropy on the Galactic magnetic field} \label{sec:results}
% \paragraph{\textbf{Model and methods:}}
We follow the analysis of~\citet{BF23}, which refined and extended the work of~\citet{ding_imprint_2021} where the source distribution follows the extragalactic matter distribution based on CosmicFlows 2~\citep{Hoffman_2018} within 350 Mpc and
a uniform source distribution is assumed for larger distances~\footnote{We checked that varying the source evolution outside of the CosmicFlows volume from $(1+z)^0$ to $(1+z)^{\pm 3}$ leads to relative differences of the predicted dipole amplitude $\leq \pm 20\%$, i.e. smaller than those from varying the coherent model of the GMF within the \texttt{UF23} suite, see Fig.~\ref{fig:dip_quad}.}. The UHECR emission spectrum was fitted to the measured cosmic ray energy spectrum~\citep{Auger_spectrum_2020} and composition~\citep{a_yushkov_for_the_pierre_auger_collaboration_mass_2019} at Earth, and the sources were assumed to be identical.
For calculating the Galactic magnetic field deflections, we use the eight new \texttt{UF23} models of the coherent field~\citep{UF23} and adopt the Planck re-tune~\citep{Planck_2016} of the \texttt{JF12} random field~\citep{jansson_galactic_2012} (``Planck" in the following, but ``Pl" in figures) as the baseline choice.  %We investigate the impact of the uncertainty in the coherent field, represented by the diversity within the \texttt{UF23} model suite.
The Planck-tuned random field scales down the amplitude of the \texttt{JF12} random field, as well as refits some other parameters like the amplitude of the Perseus spiral arm, to take into account the improved component separation in WMAP7 and Planck with respect to WMAP5 upon which \texttt{JF12} was based. We take $l_c=60\,\mathrm{pc}$ as a benchmark coherence length and evaluate the influence of ``Galactic variance" from the particular realization of the random field by using a second realization of that field. Additionally, we consider a model with $l_c=30\,\mathrm{pc}$.
As a further set of comparisons, we also show predictions for the two \texttt{JF12} models used by~\citet{BF23}: the \texttt{JF12} model with the original \texttt{JF12} random field~\citep{jansson_galactic_2012, jansson_new_2012} (\texttt{JF12-full}), and the solely coherent version (\texttt{JF12-reg}). To be able to judge the effect of the Planck-tuned random field against the original \texttt{JF12} random field, we also show \texttt{JF12} with the two realizations of the Planck random field with $l_c=60\,\mathrm{pc}$ (\texttt{JF12-Pl}).\\

% \subsection{Dipole direction} \label{sec:dipole_direc}
{\bf \noindent Dipole direction}\\
\begin{figure*}[ht]
\subfloat[energy $>8\,\mathrm{EeV}$]{\includegraphics[height=4cm]{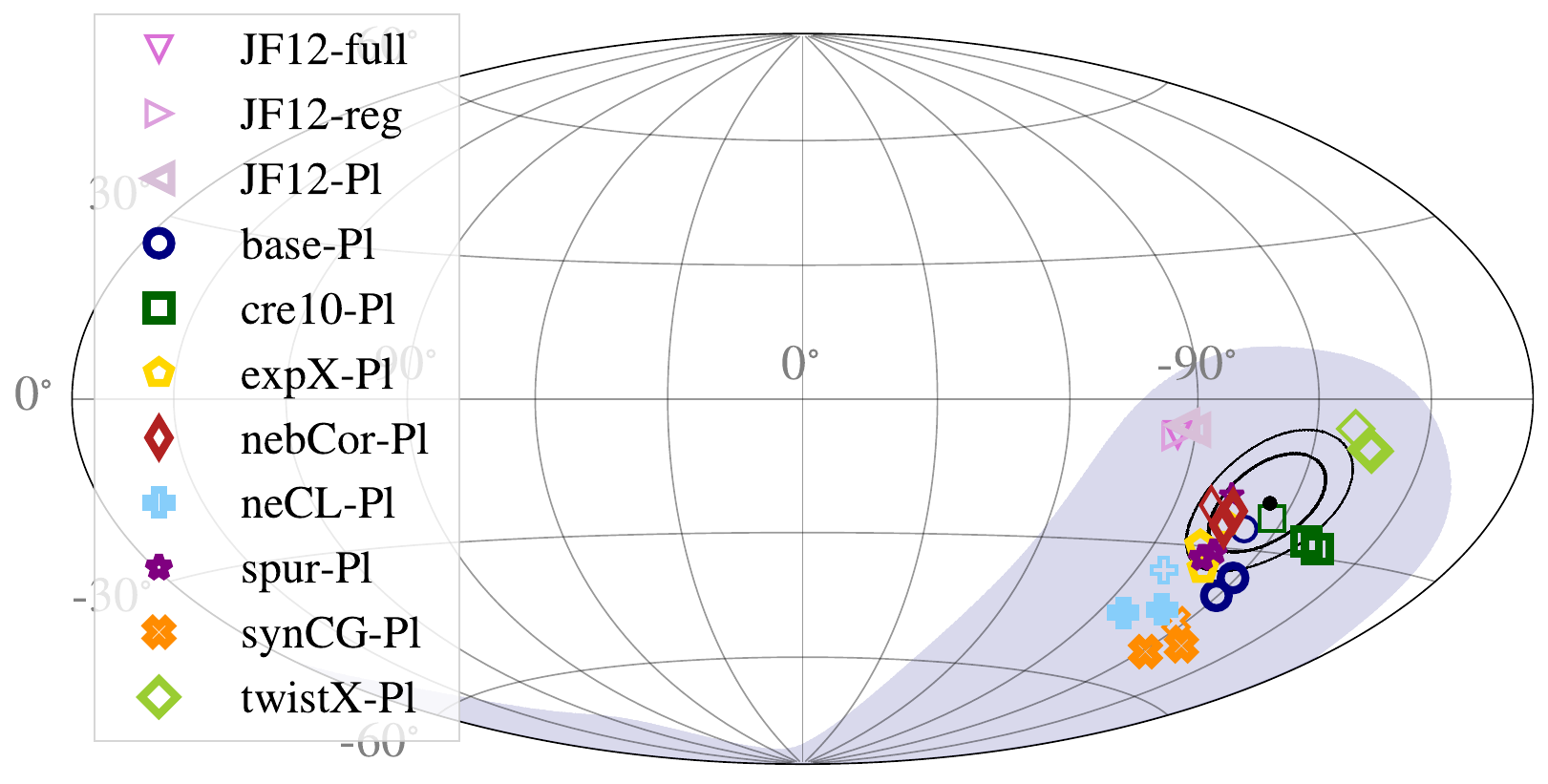}}
\subfloat[$(8-16)\,\mathrm{EeV}$]{\includegraphics[trim={16.1cm 0 0 0}, clip, height=4cm]{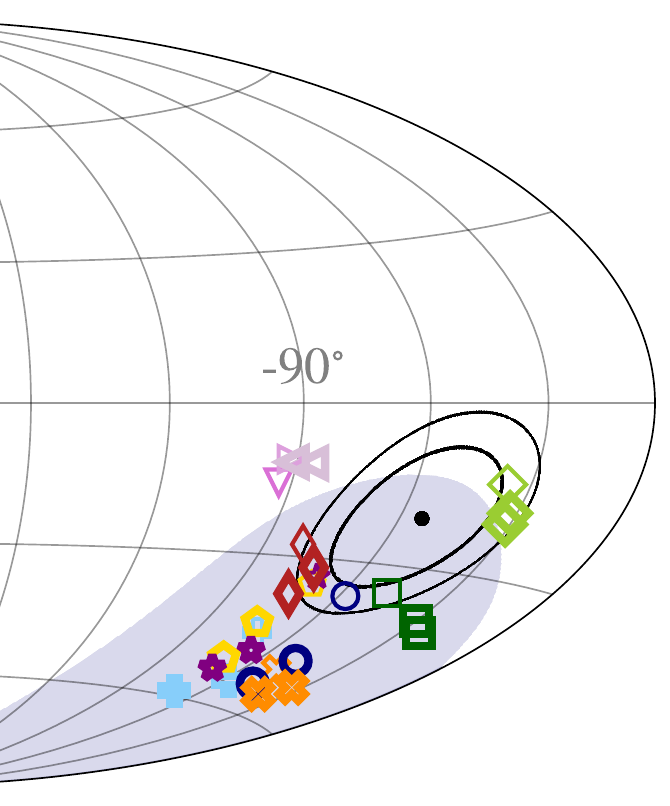}}
\subfloat[$(16-32)\,\mathrm{EeV}$]{\includegraphics[trim={16.1cm 0 0 0}, clip, height=4cm]{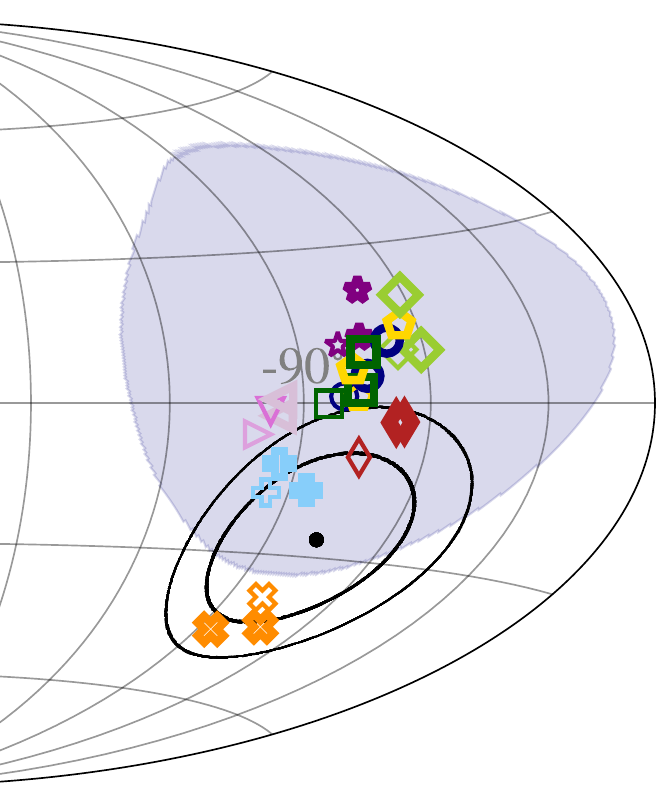}}
\subfloat[$>32\,\mathrm{EeV}$]{\includegraphics[trim={16.1cm 0 0 0}, clip, height=4cm]{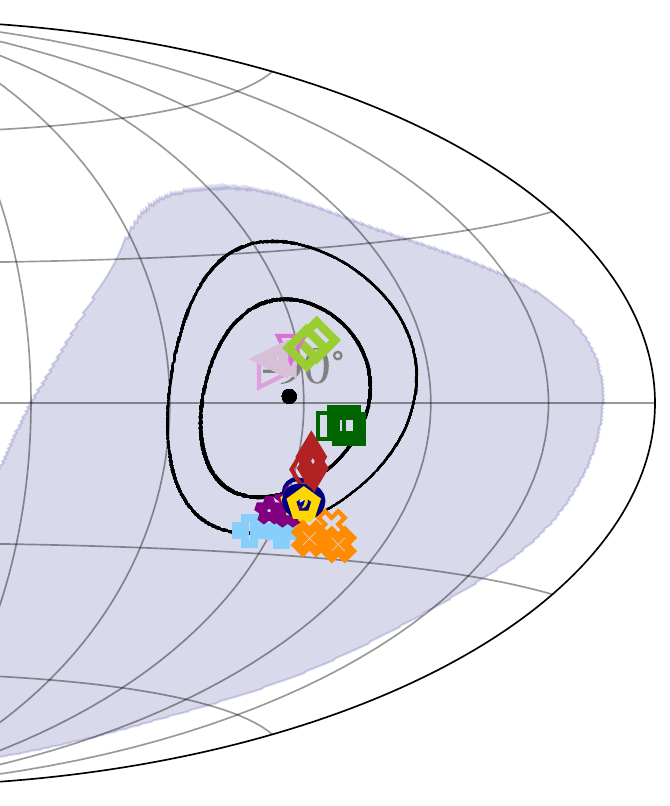}}
\caption{Predicted and measured dipole directions in Galactic coordinates:
Colored markers indicate the dipole directions for different coherent GMF models and two realizations of the random field with $l_c=60\,\mathrm{pc}$ (duplicate heavy symbols) and one with $l_c=30\,\mathrm{pc}$ (light symbols). The blue region shows the $1\sigma$ uncertainty due to cosmic variance in the source positions, for the \texttt{base} model with $l_c=60\,\mathrm{pc}$ and $n_s=10^{-3}\,\mathrm{Mpc}^{-3}$. The black contours represent the $1\sigma$ and $2\sigma$ uncertainty domains of the measured dipole~\citep{Golup_ADs_2023}.}
\label{fig:dipole_direcs}
\end{figure*}
In Fig.~\ref{fig:dipole_direcs}, the predicted directions of the dipole are shown for the 8 different \texttt{UF23} models, as well as for the three tested \texttt{JF12} models for different energy intervals. The dipole direction varies between energy bins for each GMF model despite the relatively constant rigidity. This is due to the decreasing propagation length with the energy, which leads to a variation of the distribution of contributing sources between energy bins; see also~\citet{BF23}. The \texttt{UF23} models differ by up to $\sim50^\circ$ from each other, considering differences over all energy bins. In general, all models predict the dipole direction relatively close to the indicated $1\sigma$ and $2\sigma$ contours of the measured dipole direction, consistent with the origin of the dipole being the anisotropic extragalactic source distribution following the LSS.
The dipole directions of the \texttt{UF23} models are in general more south and further away from the Galactic center than found with the \texttt{JF12} models, especially for lower energies. Additionally, it is visible that the differences between the two random field realizations with $l_c=60\,\mathrm{pc}$ and the two tested coherence lengths ($l_c=30\,\mathrm{pc}$ and $l_c=60\,\mathrm{pc}$, see above) are around $\mathcal{O}(15^\circ)$ at lower energies and $\mathcal{O}(5^\circ)$ for $E>32\,\mathrm{EeV}$ and are thus subdominant to the differences between models.
Because the dipole direction hardly differs between the three \texttt{JF12} models with entirely different random fields (Planck random field, \texttt{JF12} random field and no random field), the dipole directions of the \texttt{UF23} models will probably be reasonably stable in regard to updates of the random field model.

The dipole direction depicted in Fig.~\ref{fig:dipole_direcs} is calculated from the model for the idealized continuum case of infinite source number density\footnote{In this publication, $n_s$ is the density of \textit{contributing} sources. It may be smaller than the actual density of sources if the source emission is strongly beamed~\citep{farrar2024binary}. For transient sources, it is $n_s \simeq \Gamma \tau$ where $\Gamma$ is the volumetric rate of the transients and $\tau$ is the mean arrival time spread, which depends on the extragalactic magnetic field, see~\citet{BF23} and references therein.} $n_s=\infty$. For a more realistic treatment where sources are discrete and randomly distributed following the LSS, variations of the dipole direction are expected due to cosmic variance. These variations increase strongly with decreasing $n_s$. As in \citet{BF23}, we investigate the influence of a finite source density by randomly drawing $10,000$ explicit catalogs of sources from the continuous source distribution for each value of $n_s$. The regions encompassing 68\% of all dipole directions for the $10,000$ simulations are shown in blue in Fig.~\ref{fig:dipole_direcs}, for the \texttt{UF23-base} model and $n_s=10^{-3}\,\mathrm{Mpc}^{-3}$ (for reference, the density of Milky Way-like galaxies is $n_s\sim10^{-2}\,\mathrm{Mpc}^{-3}$~\citep{Conselice_2016}).
The uncertainty due to cosmic variance in source locations is significantly greater than that from the variations between the different coherent models or realizations of the random field. This means that the systematic uncertainty in reconstructing the origin of the UHECR dipole is not dominated by the uncertainty on the GMF, within variations of the \texttt{UF23} models. Even though subdominant, there are subtle differences between the individual \texttt{UF23} models regarding how well they reproduce the measured dipole direction. This is discussed further in App.~\ref{app:combination}.\\

% \subsection{Dipole and quadrupole amplitudes}
{\bf \noindent Dipole and quadrupole amplitudes}\\
In addition to the dipole direction, the dipole amplitude and its energy evolution are important observables that should be reproduced.
We choose the quadrupole moment as a representative of all higher multipoles as it is the first to be outside isotropic expectations (see Fig.~10 of \citet{BF23}), and its measured value including uncertainties is reported in~\citet{Caccianiga_ICRC2023}.  Figure~\ref{fig:dip_quad} depicts the dipole and quadrupole moments for the different \texttt{UF23} GMF models, for various source densities. As the source density decreases, cosmic variance from one realization to the next increases, greatly expanding the variation in predictions relative to the case of high source density and increasing the mean values of the dipole and quadrupole amplitudes.

It is noteworthy how similar the dipole and also quadrupole amplitudes are for all \texttt{UF23} models. The variations between different \texttt{UF23} coherent field models are nearly as small as the variation between the tested random field realizations.  The predictions of the \texttt{UF23} and the \texttt{JF12} models are also similar for the quadrupole amplitude, when the same random field model is used. While we did not explore the sensitivity to the random field model for the \texttt{UF23} suite, we did check for the three \texttt{JF12} models covering a range of random magnetic field strengths and coherence lengths that the dipole and quadrupole moments decrease almost linearly with increasing amplitude of the turbulent field part -- as expected since the random field smoothes out the structure.

However, the dipole amplitude is distinctly smaller in the continuum limit for all \texttt{UF23} models than for even the \texttt{JF12-full} model, which has the strongest random field. As elucidated below, the reason for the reduced dipole amplitude with the \texttt{UF23} models relative to the \texttt{JF12} models can be traced to the interplay between the peaks in the extragalactic flux distribution and the region of strong demagnification from the Galactic magnetic field. On account of this intricate relation, conclusions on the compatible range of source number densities are presently subject to large uncertainties and may change in the future when updated models for the random part of the Galactic magnetic field and better constraints on its coherence length become available. Other uncertainties come from the extragalactic magnetic field, which is largely unknown, and the details of the large scale structure. In the App.~\ref{app:EGMF}, we discuss the possibility of a non-negligible extragalactic magnetic field.

\begin{figure}[ht]
\centering
\includegraphics[width=0.46\textwidth]{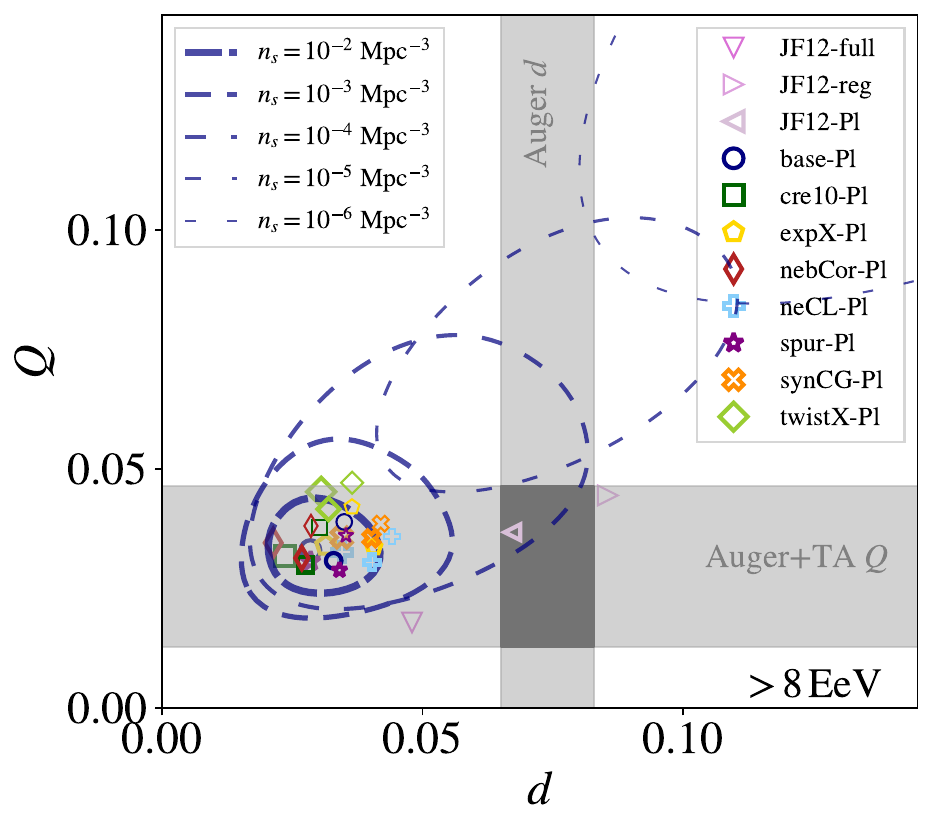}
\caption{The markers show the dipole and quadrupole moments, $d$ and $Q$, in the limit of continuous source density ($n_s=\infty$), for the various GMF models and energy $>8\,\mathrm{EeV}$ as in Fig.~\ref{fig:dipole_direcs}. The dashed curves show, for the \texttt{base} GMF model and different values of $n_s$, the $1\sigma$ domain from cosmic variance. The gray regions mark the data $1\sigma$ uncertainty for the dipole~\citep{Golup_ADs_2023} and the quadrupole~\citep{Caccianiga_ICRC2023}. Note that the measured quadrupole amplitude is not significant as it is compatible with isotropy within $2\sigma$.}
\label{fig:dip_quad}
\end{figure}

In Fig.~\ref{fig:dip_quad} we show how the $1\sigma$ regions of dipole and quadrupole amplitudes evolve with the effective source number density, $n_s$, for the \texttt{UF23-base} model. Details and a visualization of the dipole and quadrupole moments for all models and energy bins, including uncertainties, is given in Fig.~\ref{fig:dip_quad_edep} in App.~\ref{app:dip_quad}.
Cosmic variance leads to larger variations of the dipole and quadrupole amplitudes than the variations between the different \texttt{UF23} models already for $n_s=10^{-2}\,\mathrm{Mpc}^{-3}$. Therefore, conclusions on the source number density can be drawn with little sensitivity to the specific \texttt{UF23} model.
For large source densities, the dipole amplitude in the $>8\,\mathrm{EeV}$ energy bin is smaller than the measured one for all \texttt{UF23} models\footnote{The mean source distance decreases with the energy on account of composition evolution and energy losses during propagation (see Fig.~1 in~\citet{BF23}). Thus, the too-low model dipole amplitude at lower energy cannot be increased by a local source in a suitable direction without increasing even more the already large amplitude for $>32\,\mathrm{EeV}$ - at least if that source follows the same emission as all others~\citep{Auger_CFAD_2023}.}. Hence, contrary to the findings of~\citet{BF23} using the \texttt{JF12} model which showed compatibility for $n_s\geq10^{-3.5}\,\mathrm{Mpc}^{-3}$, the present analysis using the \texttt{UF23} models is incompatible with the continuous case and fits best for smaller number density. For the dipole and quadrupole amplitudes of the \texttt{UF23} models to be compatible with the measured ones for $E\geq 8$ EeV within $1\sigma$, in at least a fraction of realizations, the source density has to be $n_s\sim10^{-4}\,\mathrm{Mpc}^{-3}$ (see also Figs.~\ref{fig:inside} and~\ref{fig:dip_quad_edep}).
We stress that, as will be discussed in the next section, this conclusion is potentially sensitive to the random field.

\section{(De)magnification by the Galactic magnetic field} \label{sec:magnification}
\begin{figure*}[ht]
    \centering
    \subfloat[\texttt{JF12} + Planck  $l_c=60\,\mathrm{pc}$]{\includegraphics[width=5.9cm]{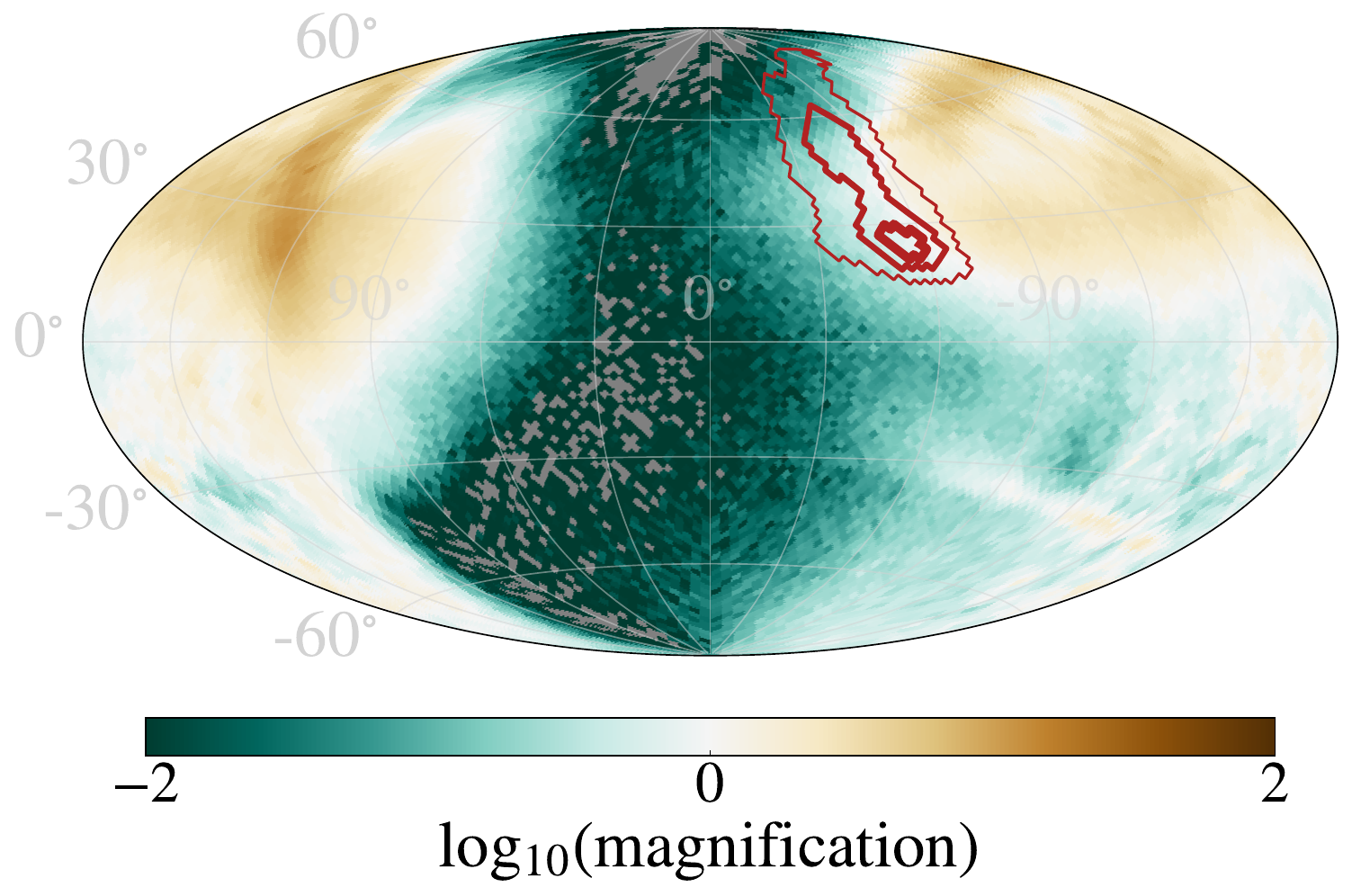}}
    \subfloat[\texttt{UF23 base} + Planck  $l_c=60\,\mathrm{pc}$]{\includegraphics[width=5.9cm]{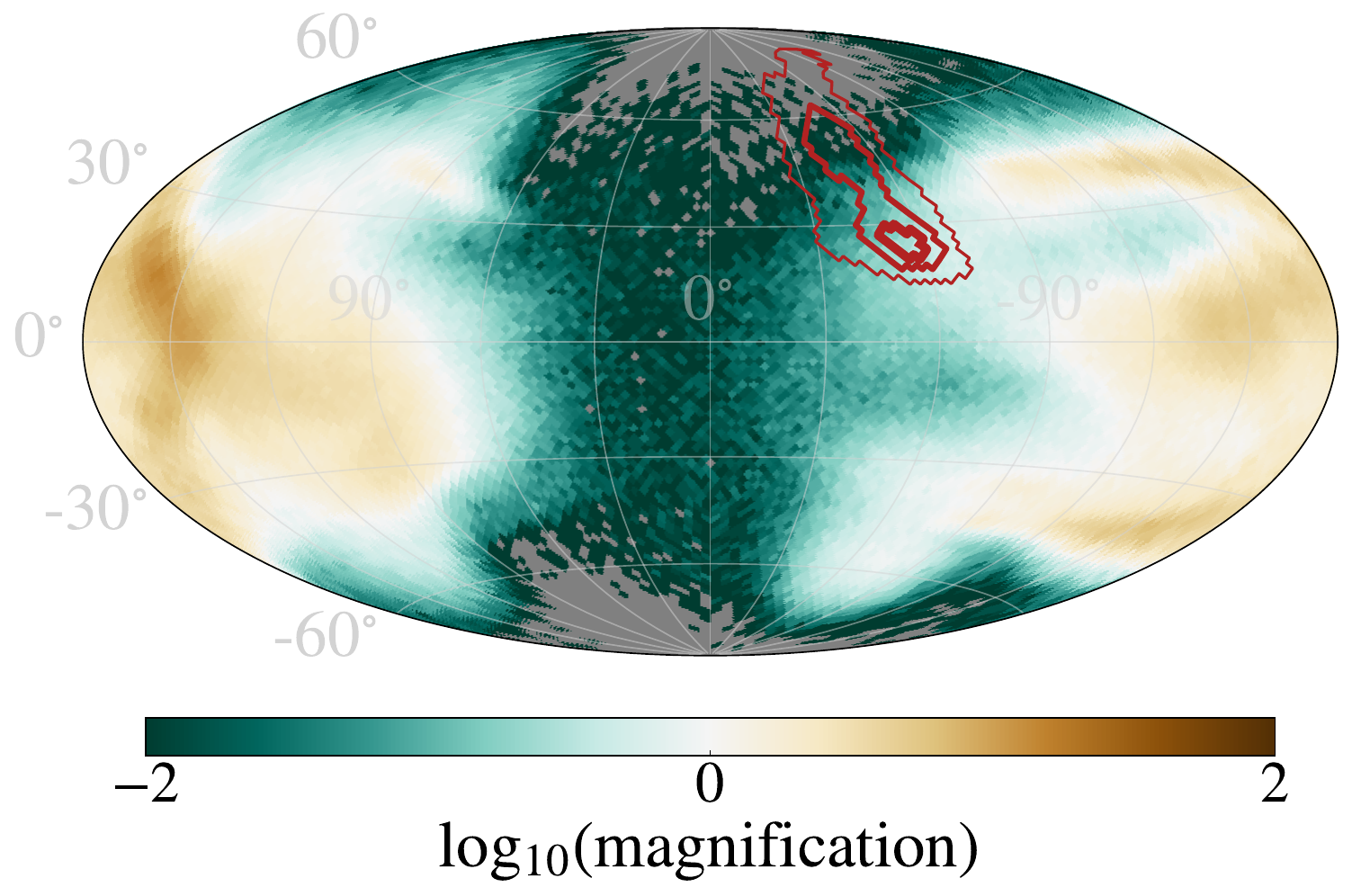}}
    \subfloat[``Illumination" map $I$\\from~\citet{BF23} \label{fig:illumination_map}]{\includegraphics[width=5.9cm]{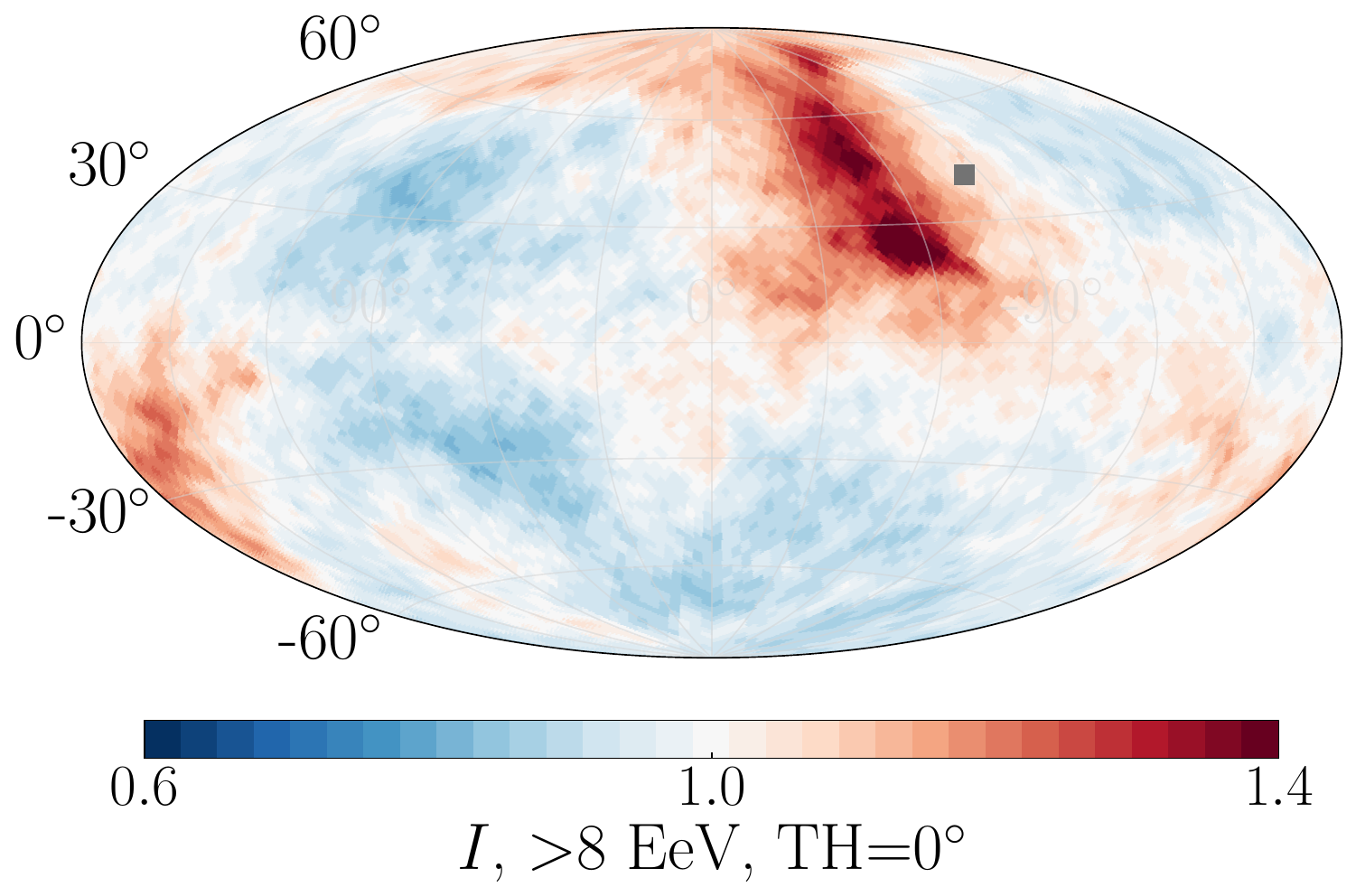}\label{fig:illum}}
    \caption{a) and b): Magnification maps for rigidity $\mathcal{R}=5\,\mathrm{EV}$ (see text for explanation). Grey pixels are source directions contributing no events at Earth.
    Contours indicating the extragalactic directions with large flux (panel (c)) are shown in red.
    c): The $E>8\,\mathrm{EeV}$ illumination map calculated from the LSS model~\citep{BF23}, showing the flux at the edge of the Galaxy.}
    \label{fig:magnification_maps}
\end{figure*}

To understand why the dipole amplitude predicted by the \texttt{UF23} models in the continuum limit is so much lower than with the \texttt{JF12} model, we show in Fig.~\ref{fig:magnification_maps} the logarithm of the \textit{magnification} for the \texttt{base} model and for the \texttt{JF12-Pl} model, for $\mathcal{R}\equiv E/eZ = 5\,\mathrm{EV}$ (the mean charge of UHECRs increases with the energy in such a way that the rigidity stays almost constant~\citep{Auger_CFAD_2023, Auger_CF_2023, Ehlert_Curious_2023} at $\mathcal{R}\approx(5\pm3)\,\mathrm{EV}$ over the whole energy range discussed here; see Fig.~4 in~\citet{BF23}). The magnification is defined to be the flux from a standard source in the respective direction, relative to the flux in the absence of the Galactic magnetic field.
Cosmic rays from some directions -- notably from sources behind the central region of the Galaxy -- are demagnified: they are deflected strongly and simply never reach the solar system. Since energy losses of UHECRs in their passage through the Galaxy are negligible, Liouville's theorem implies that the flux integrated over $4 \pi$ radians is preserved, hence the existence of demagnified directions implies directions with magnification $>1$; see~\citet{farrar_sutherland_deflections_2019} for more discussion of the mechanism and also~\citet{ Harari_2000, Harari_2002}. Corresponding maps to Fig.~\ref{fig:magnification_maps} for all \texttt{UF23} models including also variations of the random field are displayed in Fig.~\ref{fig:magnification_maps_more} in App.~\ref{app:more_magnification}.

Comparing the magnification maps to the extragalactic flux distribution according to the LSS source model (\textit{illumination}) shown in Fig.~\ref{fig:illumination_map}, one sees that the peak flux is in a demagnified region for the \texttt{UF23} models. By contrast, the \texttt{JF12} model is neutral or even magnifies the flux from those directions. The difference in magnification thus explains the significantly smaller dipole amplitude for the \texttt{UF23} models compared to the \texttt{JF12} model. Also, the invisible parts of the extragalactic illumination in the Galactic North explain why the direction of the dipole seen on Earth is displaced more towards the Galactic South for the \texttt{UF23} models.  
% That is to say, the fact that the LSS model for the extragalactic UHECR illumination is quite concentrated in the Virgo direction rather than being a smoother distribution, has a large impact on the predicted arrival directions when the (de)magnification of some directions by the GMF is taken into account.

The systematic difference between the magnification maps of the \texttt{UF23} models and \texttt{JF12} in the region of the peak of the LSS illumination map that lead to the deviation in dipole amplitude can be traced to differences in the respective toroidal halo models. This will be discussed in a separate publication.

\begin{figure*}[ht]
  \centering
\def\figw{0.32}
\includegraphics[width=\figw\linewidth]{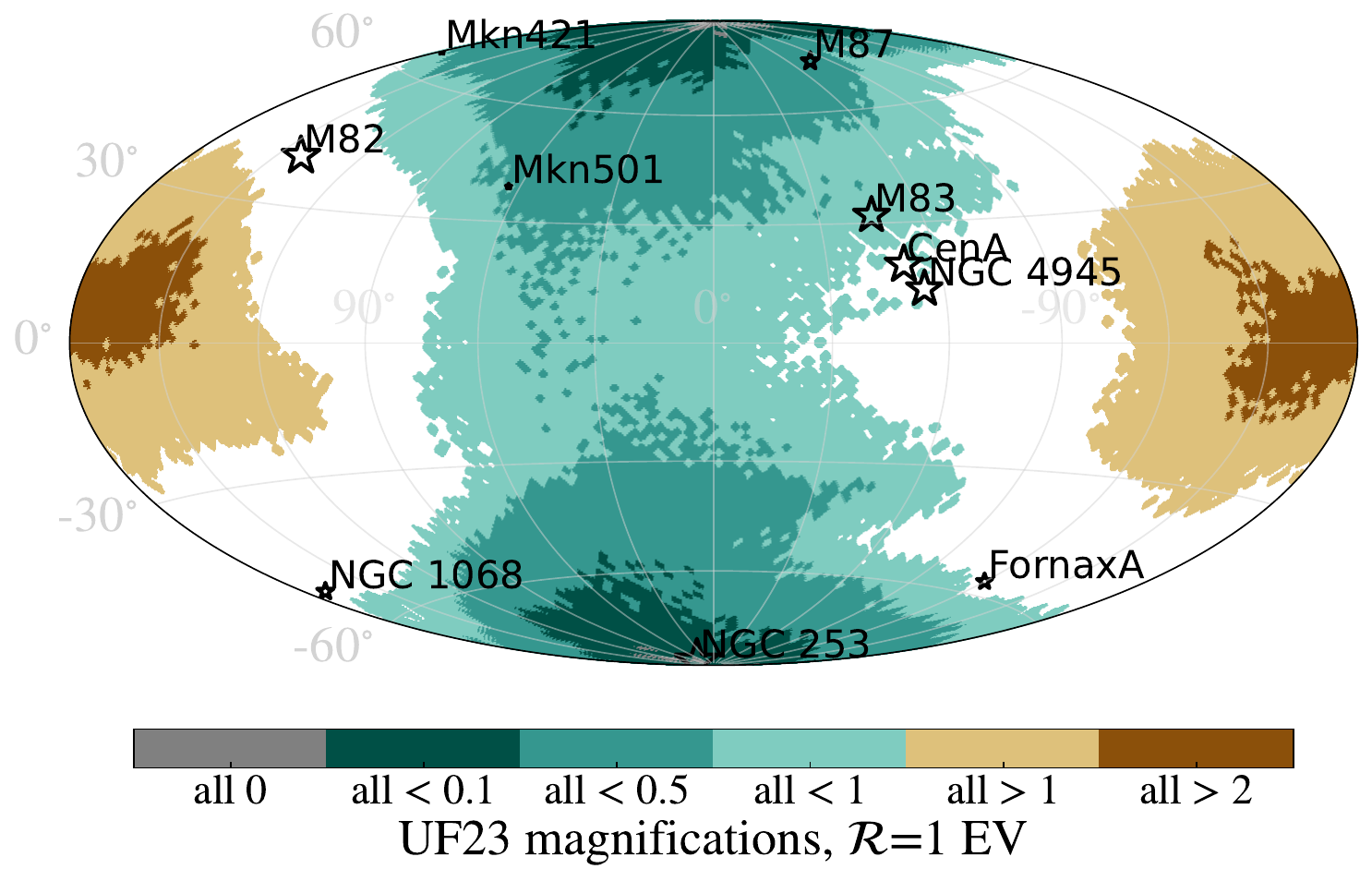}
\includegraphics[width=\figw\linewidth]{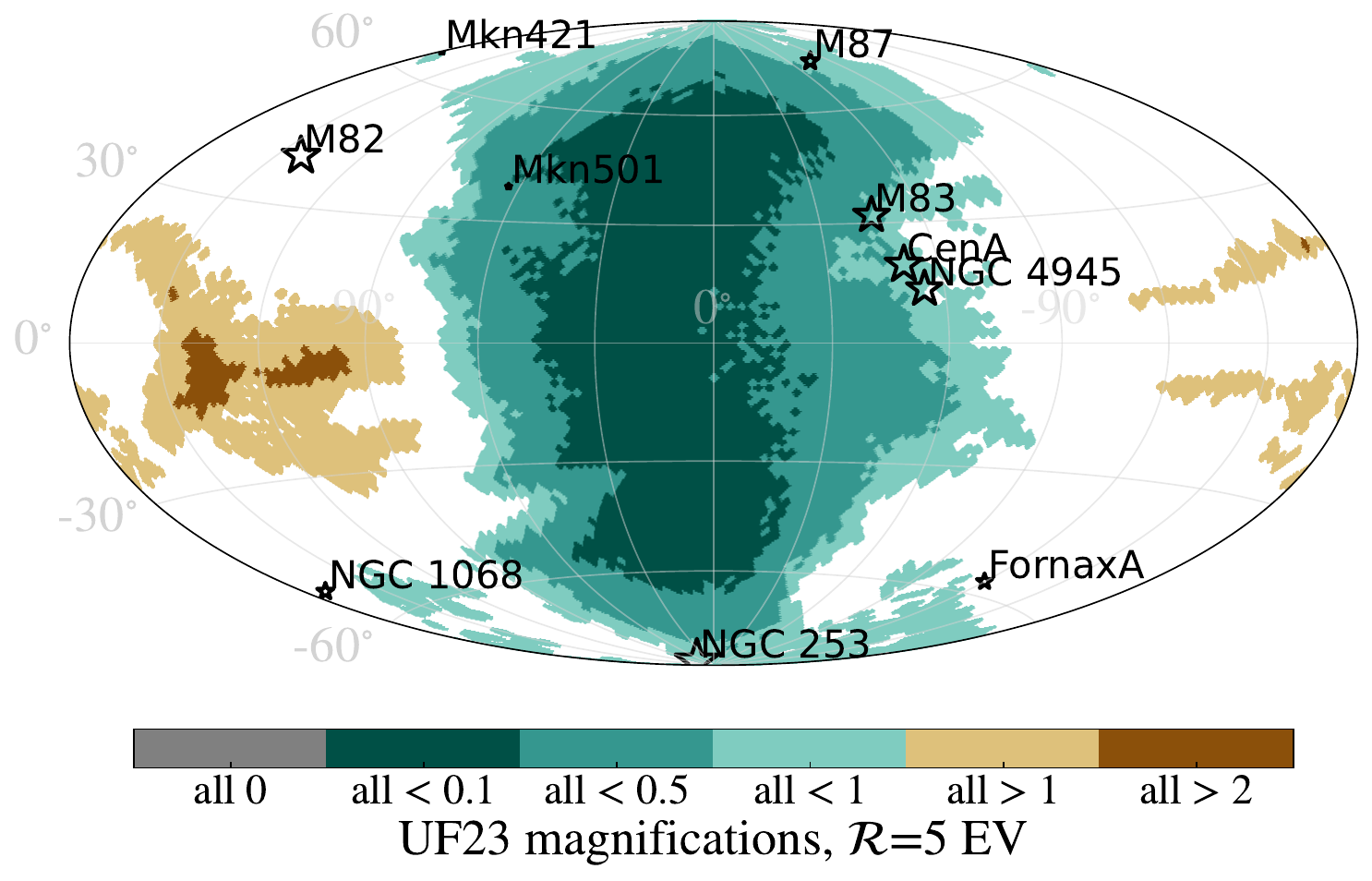}
\includegraphics[width=\figw\linewidth]{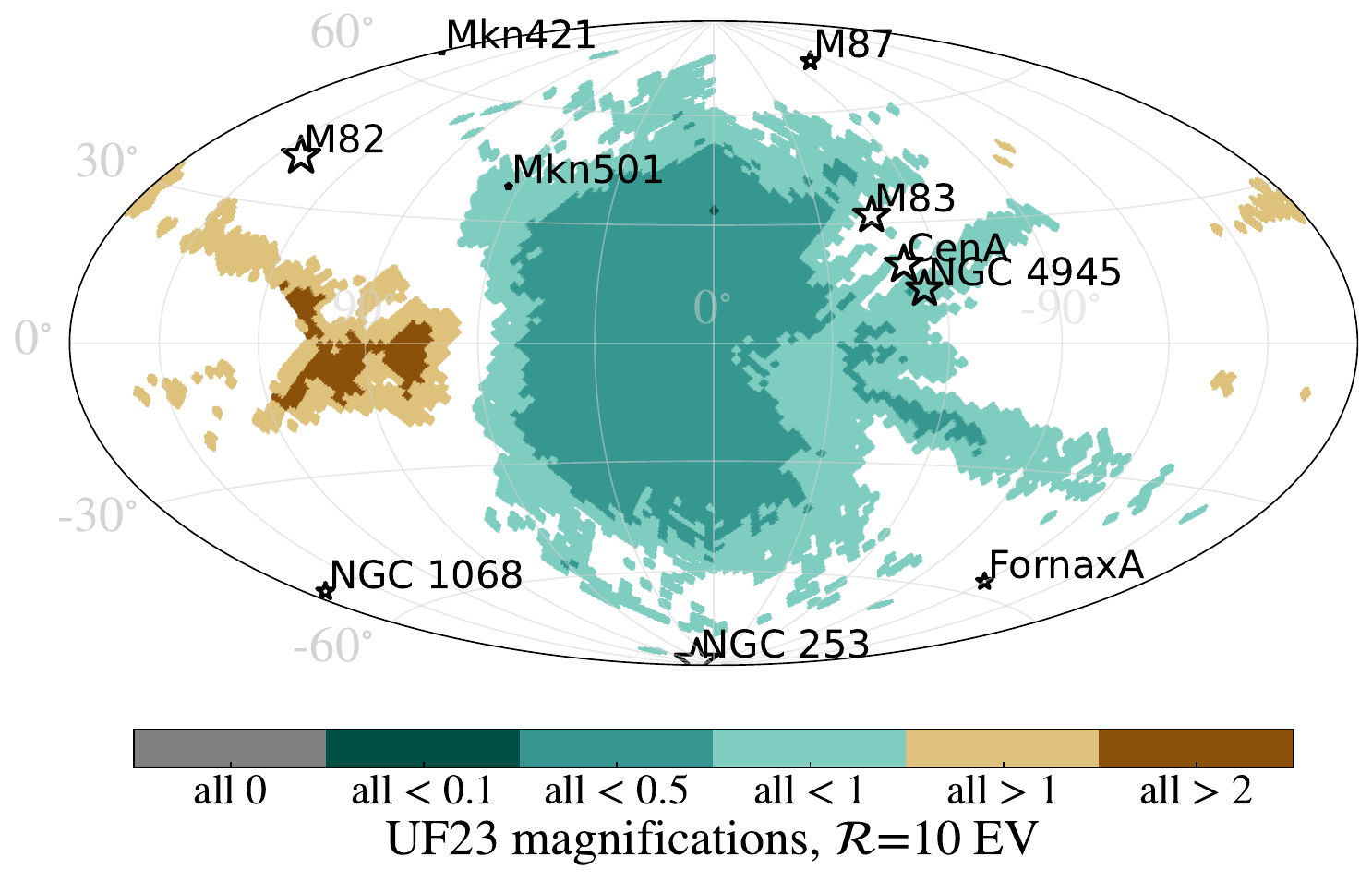}
\caption{Combined magnification maps for different rigidities of all \texttt{UF23} models with Planck random field, including both $l_c=30\,\mathrm{pc}$ and $l_c=60\,\mathrm{pc}$, the latter with two variations.
The color bar displays the magnification range in directions where all models agree; for the white area there is no consensus among the models. The directions of source candidates are indicated by stars and the marker size is proportional to $1/\mathrm{distance}$.}
\label{fig:magnification_combined}
\end{figure*}

To extract the most robust predictions for (de)magnification in the \texttt{UF23} model suite, we display in Fig.~\ref{fig:magnification_combined} the combined magnification maps for all of the \texttt{UF23} coherent and random field models studied, where the colored regions are regions of unanimity among the models and in the white regions there is no consensus (for comparison, we also show the respective maps for the \texttt{JF12} model in Fig.~\ref{fig:magnification_combined_jf12} in App.~\ref{app:more_magnification}).
It is visible that the variations between models are not very large, and that they all agree regarding the large central demagnified area. To demonstrate the implications of this, we also depict the directions of popular source candidates often used in the literature (e.g.~\citet{Auger_ADs_2022, Auger_CFAD_2023, Eichmann:2022ias, Matthews_2018}). Several of those candidates like M83, M87, Mkn421, Mkn501, Cen A, NGC4945 and NGC253 lie in the demagnification area for some or most rigidities $\mathcal{R}\lesssim5\,\mathrm{EV}$ or in some cases for all rigidities shown.

One further important consequence of the sensitivity of the dipole amplitude to the interplay between the illumination and GMF magnification maps is that the predicted dipole amplitude and direction differ substantially when the illumination is replaced by an idealized dipole with the same amplitude and direction -- for example the ``2MRS dipole" which is often used in the literature, e.g.~\citet{Auger_dipole_2017_Science, Auger_dipole_2018, Bray_2018, Bakalova_2023}. The amplitude of the predicted dipole is typically a factor-2 larger with the idealized dipole than with the LSS model for the \texttt{UF23} models, and its direction differs by $\mathcal{O}(20^\circ-60^\circ)$; see Figs.~\ref{fig:dipoleI_amp} and~\ref{fig:dipoleI} in App.~\ref{app:combination}. 
Thus, it is crucial to take into account the concentrated inhomogeneities in the UHECR arrival flux %that are produced by energy losses limiting the distance of contributing sources -- 
instead of using a smooth dipole approximation for the illumination map that is based on averaging the galaxy distribution over distances much larger than that of contributing sources.
% Since energy losses in propagation create inhomogeneities in the UHECR arrival flux (as captured at least approximately in the LSS model), predictions for the UHECR arrival direction anisotropy using a smooth dipole approximation for the illumination map based on the very large-scale dipole in the galaxy distribution averaged over distances much larger than that of contributing UHECR sources, can be expected to be misleading.

\section{Conclusions}
We have investigated the sensitivity of the predicted large-scale anisotropy of ultra-high-energy cosmic rays to the coherent part of the Galactic magnetic field, and also made a preliminary study of the dependence on the random part of the field. We find that the measured dipole which has been detected in the UHECR arrival flux~\citep{Auger_dipole_2017_Science, Auger_dipole_2018} can be described reasonably well by a model where the sources follow the large scale structure of the universe and UHECRs are deflected by any of the suite of \texttt{UF23} GMF models~\citep{UF23} (as is also the case for the \texttt{JF12} model~\citep{BF23}). For the \texttt{UF23} models, the best agreement with both dipole direction as well as dipole and quadrupole amplitudes is reached for source number densities of $\mathcal{O}(n_s)=10^{-3}\,\mathrm{Mpc}^{-3}$ in the case of negligible extragalactic magnetic field.

The variations among the predicted dipole and quadrupole amplitudes, and among the dipole directions, when using different \texttt{UF23} models including different setups of the random field are small and subdominant to cosmic variance from random source positioning. Hence, the uncertainty on UHECR arrival directions from the Galactic magnetic field modeling, within the \texttt{UF23} family, will likely not obstruct conclusions about the sources of UHECRs based on their large-scale anisotropies. At the same time, we find enough differences among models, such that in the future with refined treatment of the random field, composition sensitivity, and LSS source modeling, it should be possible to disfavor or prefer some of the models.

An important discovery of our work, which goes beyond specific models of the GMF, is the unanticipated sensitivity of the dipole {\it amplitude} to the coherent field model. This results from the delicate interplay between demagnification of flux from sources behind the central portion of the galaxy and the direction of strongest extragalactic illumination from the Virgo cluster and Great Attractor. This sensitivity of the dipole amplitude will be a powerful tool to probe not only the GMF, but also the UHECR source distribution and potentially even hadronic interaction models which impact the charge assignment. The pattern of extragalactic illumination changes with UHECR energy, which should help in discriminating different contributing factors in the future.

The area of the sky where the flux is severely demagnified in the \texttt{UF23} model suite includes popular source candidates like M87, M83 and NGC 253, which are thus not expected to contribute many UHECRs at Earth, except for rigidities $\mathcal{R} >5\,\mathrm{EV}$; see Fig.~\ref{fig:magnification_combined}. Another consequence of the demagnification is that using an idealized extragalactic dipole with the same direction and strength, but neglecting the intermediate-scale anisotropy due to energy losses, gives misleading results.

The delicate relationship between the direction and amplitude of the peak extragalactic flux and the blind directions resulting from GMF demagnification implies that conclusions about the GMF model and source number density are sensitive to details of the source distribution as well as the random part of the Galactic magnetic field and the possible influence of the extragalactic magnetic field.  Thus conclusions about the UHECR source density and the relevance of cosmic variance in the source distribution must be left to the future when these aspects of the problem are better understood.

\section*{Acknowledgements}
We thank Foteini Oikonomou for useful feedback on our analysis. The work of T.B.\ is supported by a Radboud Excellence fellowship from Radboud University in Nijmegen, Netherlands and that of G.R.F.\ is supported by National Science Foundation Grant No.~PHY-2013199. T.B.\ and M.U.\ thank the Center for Cosmology and Particle Physics of New York University for its kind hospitality and T.B.\ acknowledges the support from the Alfred P. Sloan Foundation facilitating this research and the production of this paper.

% \software{CRPropa3 \citep{alves_batista_crpropa_2016},
%           healpy \citep{healpy},  
%           numpy \citep{numpy}, 
%           scipy \citep{scipy},
%           astrotools (\href{https://git.rwth-aachen.de/astro/astrotools}{Bister et al.})
%           }



\appendix

\section{Agreement of the \texttt{UF23} models with measured dipole directions and multipole amplitudes} \label{app:combination}
\begin{figure}[ht]
\includegraphics[height=8cm]{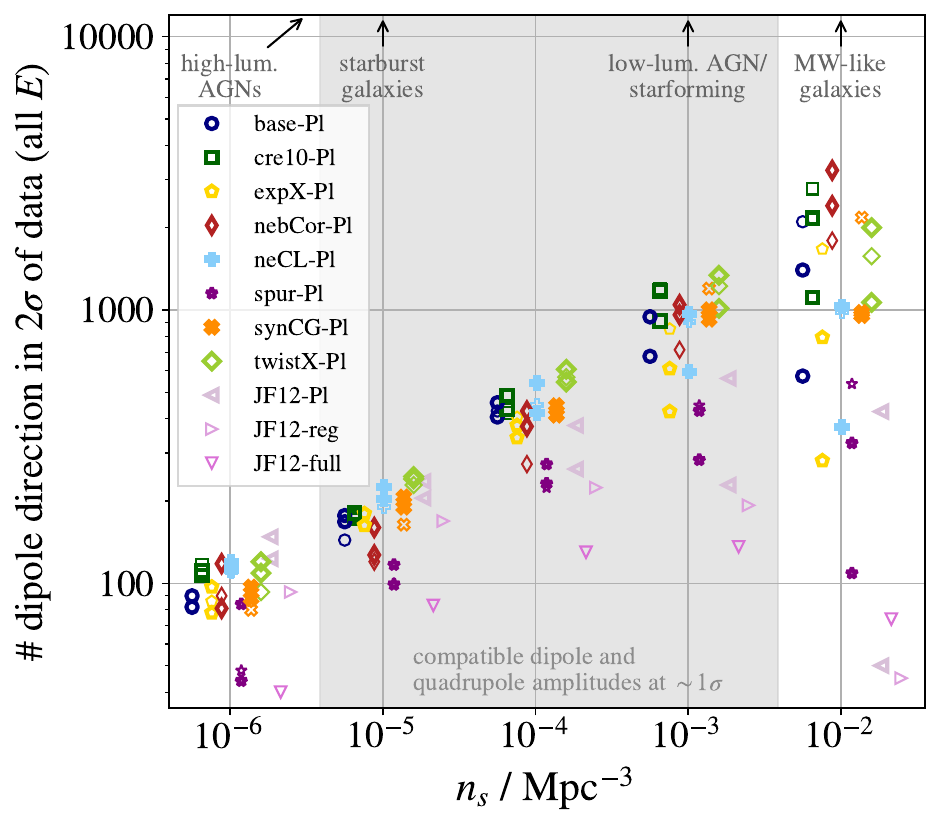}
\includegraphics[height=8cm]{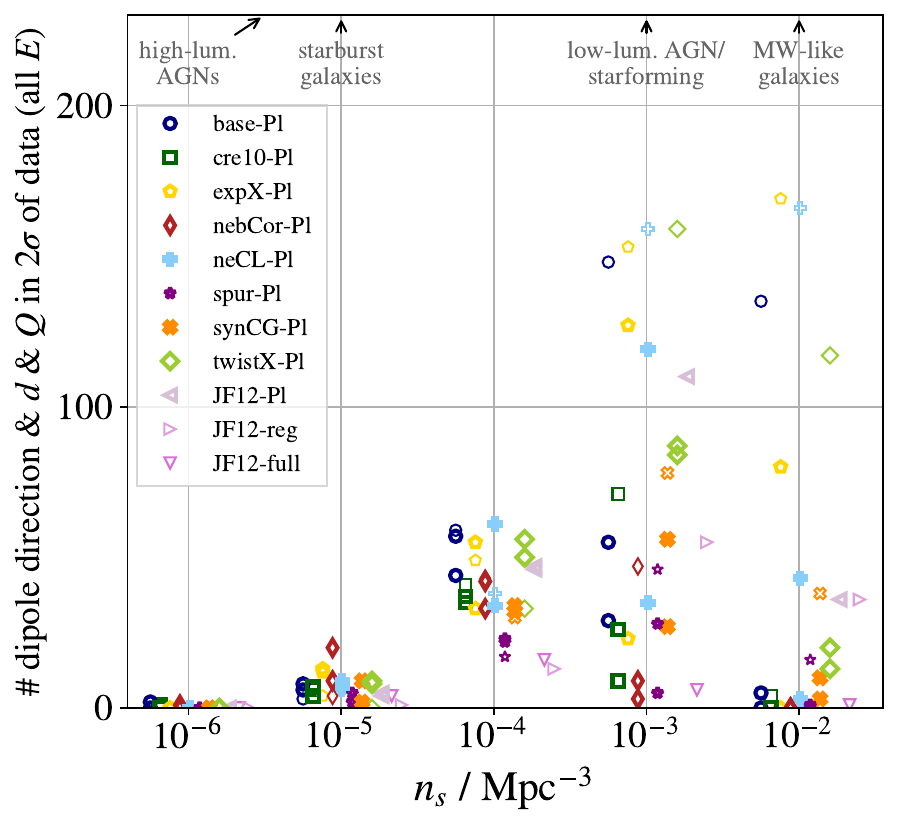}
\caption{Number of random realizations of source locations, out of $10,000$ total for each $n_s$, for which the predicted dipole direction (\textit{left}) or the dipole direction, dipole amplitude, and quadrupole amplitude all simultaneously (\textit{right}) lie within the measured $2\sigma$ uncertainty in \textit{all} of the the mutually exclusive energy bins ($E=(8-16)\,\mathrm{EeV}$, $E=(16-32)\,\mathrm{EeV}$, and $E>32\,\mathrm{EeV}$). Different GMF models are shown with different markers; their thickness indicates the random field coherence length as in Fig.~\ref{fig:dipole_direcs}. Duplicate markers denote two realizations of the turbulent field for the same coherent field. The markers for each value of $n_s/\mathrm{Mpc}^{-3}=10^{-6}, 10^{-5}, ..., 10^{-2}$ are offset on the x-axis for better visibility. The gray region (\textit{left}) marks the range of $n_s$ satisfying the dipole and quadrupole amplitudes discussed in the main text. For comparison, the arrows indicate estimates for the source number densities of different steady source candidate classes: Milky-Way-like galaxies~\citep{Conselice_2016}, low-luminosity AGNs~\citep{Ho_2008}, starforming galaxies~\citep{Gruppioni_2013}, starburst galaxies~\citep{Gruppioni_2013, Murase_2019} and high-luminosity AGNs~\citep{Gruppioni_2013, Best_2012}.}
\label{fig:inside}
\end{figure}

Even when we cannot make strong judgments regarding the quality of the dipole direction prediction for the different models due to the large influence of cosmic variance and present uncertainties on the random field and the LSS distribution, it is still possible to determine which models lead to a better agreement with the data than others.
Fig.~\ref{fig:inside} (\textit{left}) shows the number of realizations of source locations (out of $10,000$ total, for each $n_s$) for which the predicted dipole direction is within the $2\sigma$ uncertainty of the measured dipole direction, simultaneously in all mutually exclusive energy bins $E=(8-16)\,\mathrm{EeV}$, $E=(16-32)\,\mathrm{EeV}$, and $E>32\,\mathrm{EeV}$.
At large source number density, when cosmic variance between realizations is subdominant to the differences between models, there are substantial variations between the different \texttt{UF23} models and between the different random field realizations. This can be compared to Fig.~\ref{fig:dipole_direcs} which shows the dipole directions in the continuous limit.

With decreasing source density, the number of compatible realizations decreases as expected since the variance of dipole directions increases. For source densities $n_s\lesssim 10^{-3}\,\mathrm{Mpc}^{-3}$ (which gives better agreement with the measured dipole and quadrupole amplitudes than densities $n_s\gtrsim 10^{-2}\,\mathrm{Mpc}^{-3}$ for the \texttt{UF23} models as shown in Fig.~\ref{fig:dip_quad}), the differences between the models decrease and it is clearly visible that the bulk of \texttt{UF23} models gives a better fit to the dipole direction than the predecessor \texttt{JF12} model. Comparing the eight \texttt{UF23} models, no best model can be unambigously identified due to the relatively large fluctuations between different random field setups (large Galactic variance) and the variation of the number of compatible simulations with the source number density, but it can be seen that the \texttt{spur} model consistently presents a worse fit to the dipole direction than the other models.

We show in Fig.~\ref{fig:inside} (\textit{right}) the number of simulations out of $10,000$ total where the dipole direction \textit{and} the dipole amplitude \textit{and} the quadrupole amplitude are all within $2\sigma$ of the measured value. From Fig.~\ref{fig:inside} (\textit{left}) it is clear that because of cosmic variance, the number of simulations agreeing with the measured dipole direction decreases strongly with the source density. This is as expected, and thus does not necessarily exclude small source densities. But, from Fig.~\ref{fig:dip_quad} we know that densities $n_s\lesssim10^{-5}\,\mathrm{Mpc}^{-3}$ are disfavored (for negligble extragalactic magnetic field) as they lead to too large quadrupole amplitudes. Large source densities $n_s\gtrsim10^{-2}\,\mathrm{Mpc}^{-3}$ on the other hand generally lead to too small dipole amplitudes for the \texttt{UF23} models. In general, agreement with all three observables is most often reached for $n_s=10^{-3}\,\mathrm{Mpc}^{-3}$, and especially for the \texttt{twistX}, \texttt{expX}, \texttt{neCl} and \texttt{base} models. These models can even reach a compatible dipole amplitude for large source densities $n_s=10^{-2}\,\mathrm{Mpc}^{-3}$. The random field realization plays a very large role in that case, and a compatibility with all three observables is only reached for few specific models.
The \texttt{nebCor}, \texttt{synCG}, and \texttt{cre10} models, even though being a fair fit for the dipole direction as visible in Fig.~\ref{fig:inside} (\textit{left}), are less often compatible with the multipole amplitudes than the other models and are thus not favored according to Fig.~\ref{fig:inside} (\textit{right}). The \texttt{spur} model is also disfavored according to Fig.~\ref{fig:inside} (\textit{right}), but that is due to the dipole direction not fitting as well as explained above.

\section{Influence of the extragalactic magnetic field}  \label{app:EGMF}
Another important impact on the anisotropy of the UHECR flux comes from the extragalactic magnetic field (EGMF), which can dampen the multipole moments significantly. \citet{BF23} modeled this effect by a smoothing of the arrival flux of the following form~\citep{Achterberg:1999vr}:
\begin{equation}
\label{eq:delthet}
    \delta \theta = 2.9^\circ \frac{B}{\mathrm{nG}} \frac{10\,\mathrm{EV}}{E/Z} \frac{\sqrt{\overline{D} \ L_{c}}}{\mathrm{Mpc}} = 2.9^\circ \beta_{\rm EGMF} \frac{10\,\mathrm{EV}}{E/Z} \sqrt{\frac{\overline{D}}{\mathrm{Mpc}}},
\end{equation}
with the EGMF field strength $B$, the EGMF coherence length $L_c$, and the mean source distance $\overline{D}$. In the second equality, the combination $\beta_{\rm EGMF} \equiv B / \mathrm{nG}\,\sqrt{L_c / \mathrm{Mpc}}$ is introduced to isolate the quantity that can be constrained.

Using the \texttt{JF12} model for the GMF,  it was investigated by \citet{BF23} which combination of the EGMF parameter $\beta_{\rm EGMF}$ and the source density $n_s$ can produce a large enough dipole moment while keeping all higher multipole moments small enough to agree with the data. Here, we update these findings using now the \texttt{UF23} models for the GMF. For comparibility we use the same definition of two criteria (note that these criteria are slightly different than the ones used above in the main text:
\begin{enumerate}
    \item The dipole moment in the Auger field of view at $E>8\,\mathrm{EeV}$ should be $d_\mathrm{8\,EeV}>5\%$. This value is around $2.5\sigma$ below the value measured by Auger~\citep{Golup_ADs_2023}, so it constitutes approximately a $99\%$ C.L. lower limit on the dipole amplitude.
    \item All the higher multipole moments $C_{l>1}$ must be within the 99\% isotropic expectation for all energy ranges ($E=(8-16)\,\mathrm{EeV}$, $E>8\,\mathrm{EeV}$, $E=(16-32)\,\mathrm{EeV}$, and $E>32\,\mathrm{EeV}$).
\end{enumerate}

\begin{figure}[ht]
\centering
\includegraphics[width=0.48\textwidth]{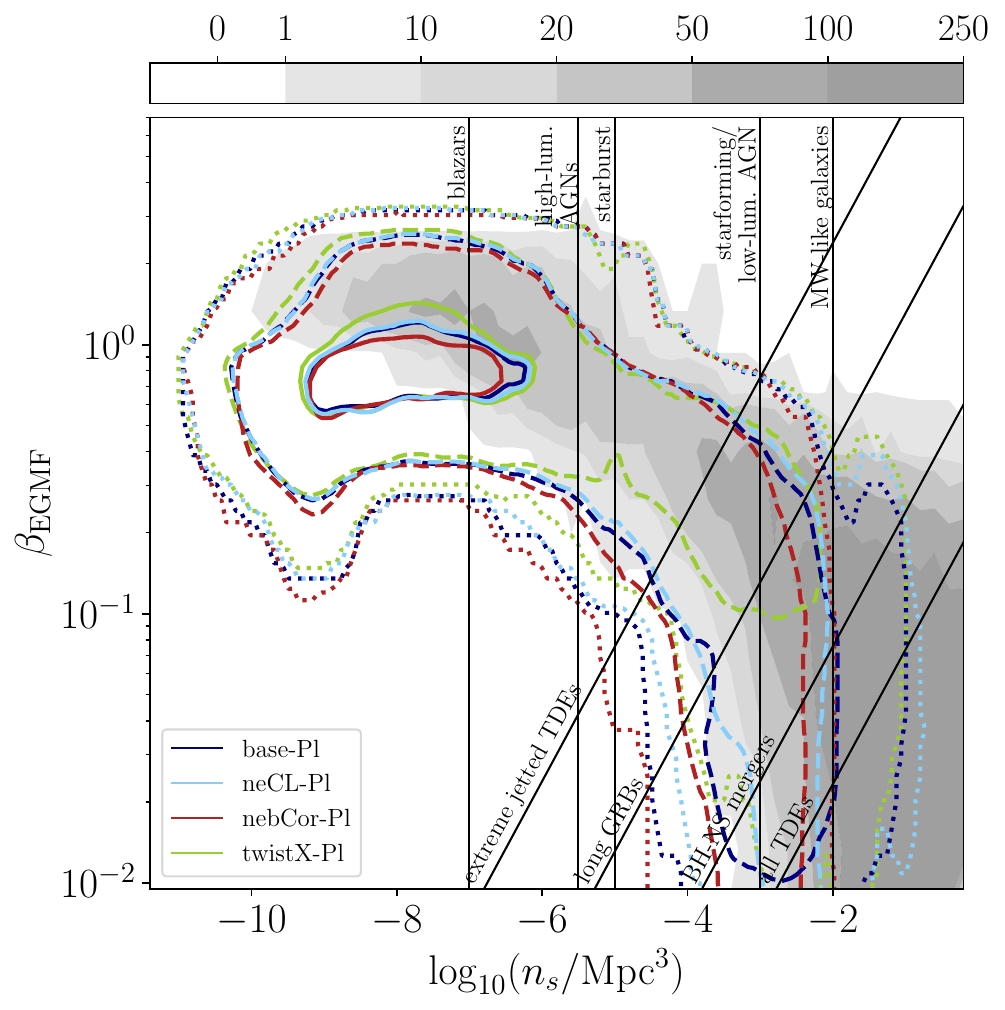}
\caption{Combined constraints on the the source number density $n_s$ and EGMF parameter $\beta_\mathrm{EGMF} \equiv B / \mathrm{nG}\,\sqrt{L_c / \mathrm{Mpc}}$. Taken from~\citet{BF23} (Fig. 11), the grey filled contours with the intensity bar at the top show the number of simulations out of 1000 total, using the  \texttt{JF12-full} model for the GMF, that have both a sufficiently large dipole and higher multipole moments small enough to be compatible with the 99\% isotropic expectations as found for the data. The navy, lightblue, red, and green contours indicate the regions encompassing 1 (dotted), 20 (dashed), and 100 (solid) simulations that fulfill both criteria for the \texttt{base}, \texttt{neCL}, \texttt{nebCor}, and \texttt{twistX} models.
Characteristic estimates of the number densities of some steady source candidates are shown with vertical lines (Milky-Way-like galaxies~\citep{Conselice_2016}, low-luminosity AGNs~\citep{Ho_2008}, starforming galaxies~\citep{Gruppioni_2013}, starburst galaxies~\citep{Gruppioni_2013, Murase_2019}, high-luminosity AGNs~\citep{Gruppioni_2013, Best_2012}, and blazars~\citep{Ajello_2013}).
Indicative locii of transient source candidates are shown with rotated lines (long GRBs~\citep{grbRatePiran10}, tidal disruption events (TDEs)~\citep{vVfTDErate14, Andreoni_2022}, and black hole - neutron star mergers~\citep{LVK_binaryMergerRate23}), see~\citet{BF23} for more details.}
\label{fig:EGMF}
\end{figure}

In Fig.~\ref{fig:EGMF}, the number of simulations that fulfill both criteria simultaneously is visualized, both for the \texttt{JF12-full} and a selection of the \texttt{UF23} models. Here we choose to show the \texttt{base}, \texttt{nebCor}, \texttt{neCL}, and \texttt{twistX} models where the latter three exhibit the smallest and largest dipole amplitudes and the largest quadrupole amplitude, respectively (Fig.~\ref{fig:dip_quad_edep}).
While for the \texttt{JF12} model negligible field strengths and very large source densities were preferred, for all shown \texttt{UF23} models smaller source densities and larger extragalactic magnetic fields are favored. This is as expected due to the smaller dipole amplitudes with the \texttt{UF23} models that have to be compensated by smaller source densities as seen above. The parameter space that is compatible with the data is very broad and extends over multiple orders of magnitude in source density. The region where most simulations fulfill both criteria stated above is at very small source densities of around $n_s\sim10^{-8}\,\mathrm{Mpc}^{-3}$ in combination with a sizeable EGMF of around $\beta_{\rm EGMF}\sim1\,\mathrm{nG}\,\mathrm{Mpc}^{1/2}$. Comparing this to the densities of different source classes as indicated in Fig.~\ref{fig:EGMF}, the favored region overlaps with the density of blazars~\citep{Ajello_2013}. More abundant source classes like starburst galaxies, other types of AGNs or even Milky-Way like galaxies are however also all compatible with the \texttt{UF23} models. The same is true for all transient source classes indicated in Fig.~\ref{fig:EGMF}. Note, as shown in sec.~\ref{sec:results}, the dipole direction is almost completely random for these small source densities.

As in the case without EGMF described above, the region of the parameter space of $n_s$ and $\beta_{\rm EGMF}$ is very sensitive to the LSS model and the random part of the Galactic magnetic field and may hence be subject to change once updated models of that become available. Additionally, a more accurate treatment of the EGMF deflections than the simplified smearing we employ here could lead to changes of the compatible values for the source density and EGMF parameters.

\section{Dipole and quadrupole amplitudes} \label{app:dip_quad}
\begin{figure*}[ht]
  \centering
  \def\figwl{0.43}
  \def\figwr{\figwl}
\includegraphics[width=\figwl\textwidth]{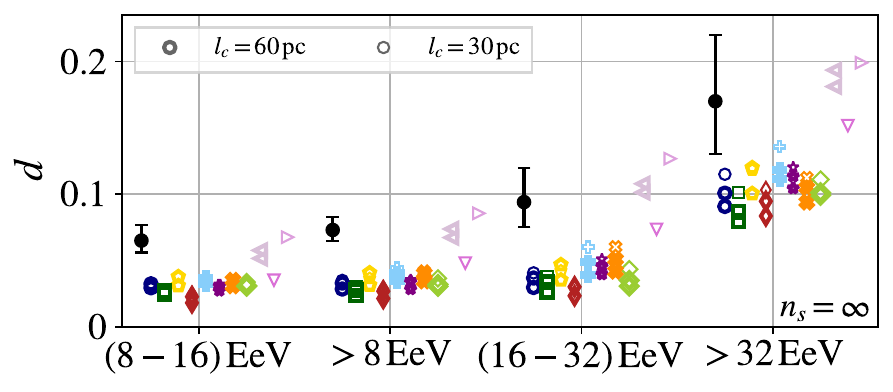} \includegraphics[width=\figwr\textwidth]{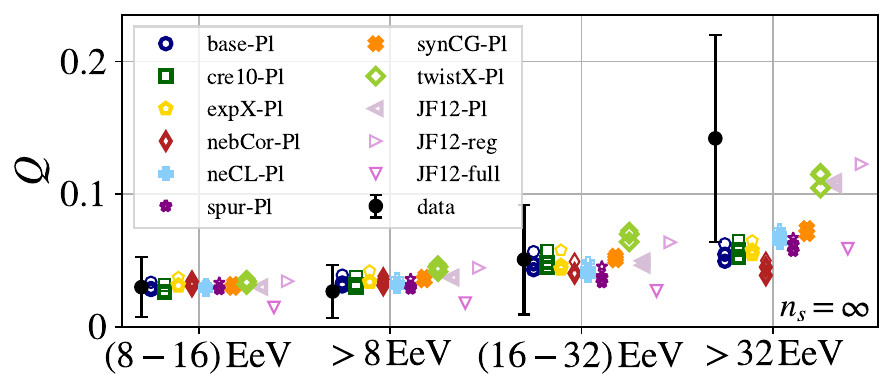}
\\
\includegraphics[width=\figwl\textwidth]{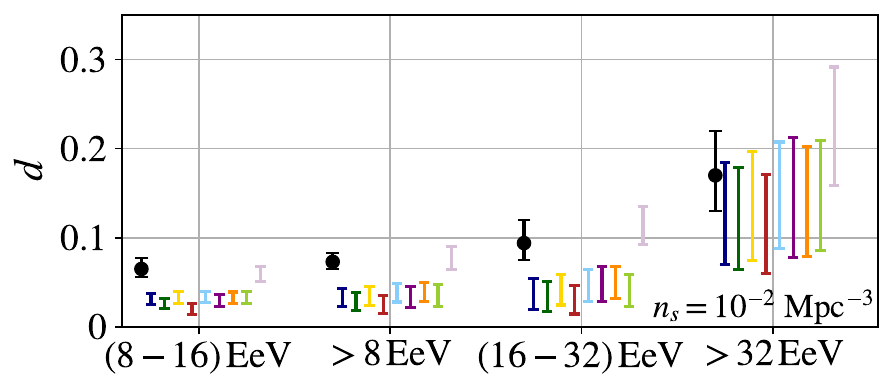} \includegraphics[width=\figwr\textwidth]{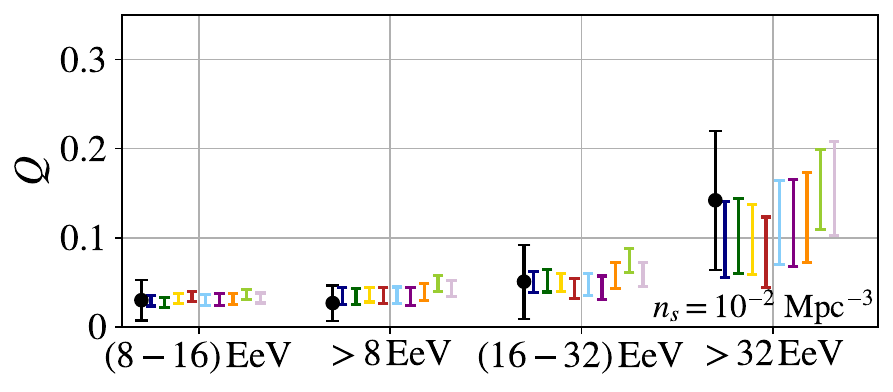}
\\
\includegraphics[width=\figwl\textwidth]{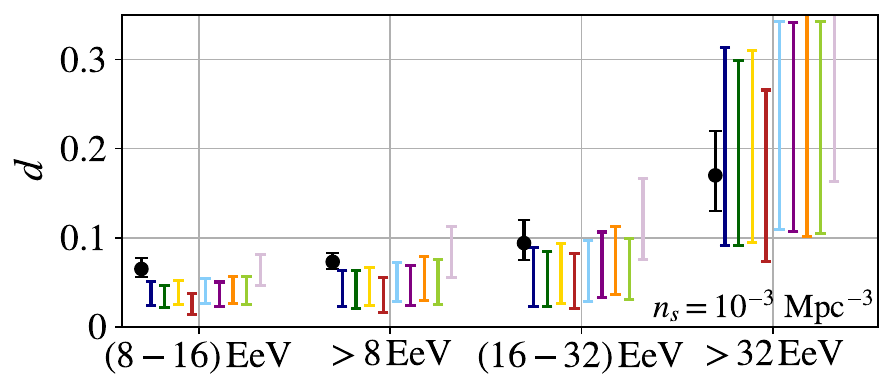} \includegraphics[width=\figwr\textwidth]{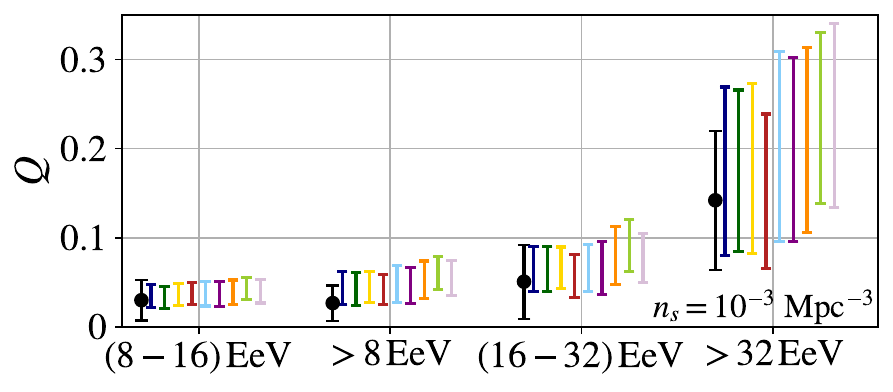}
\\
\includegraphics[width=\figwl\textwidth]{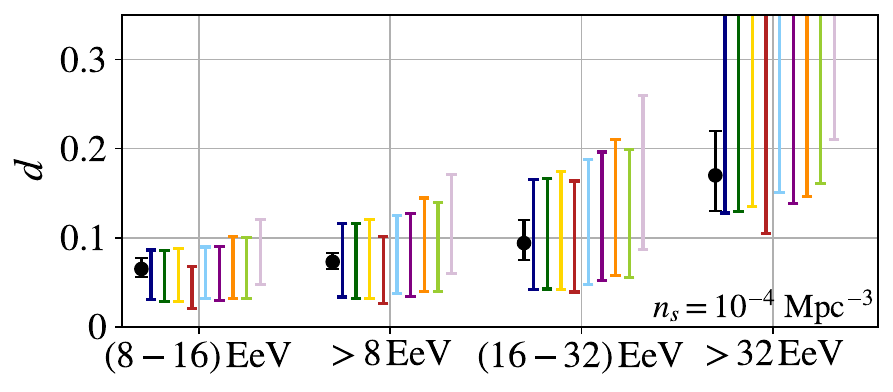} \includegraphics[width=\figwr\textwidth]{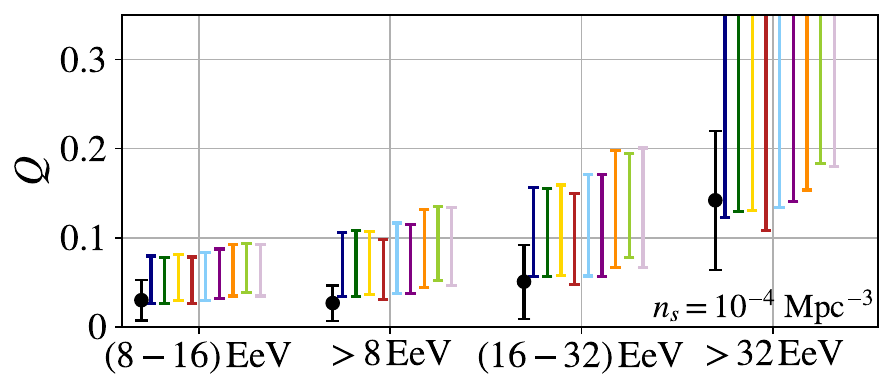}
\\
\includegraphics[width=\figwl\textwidth]{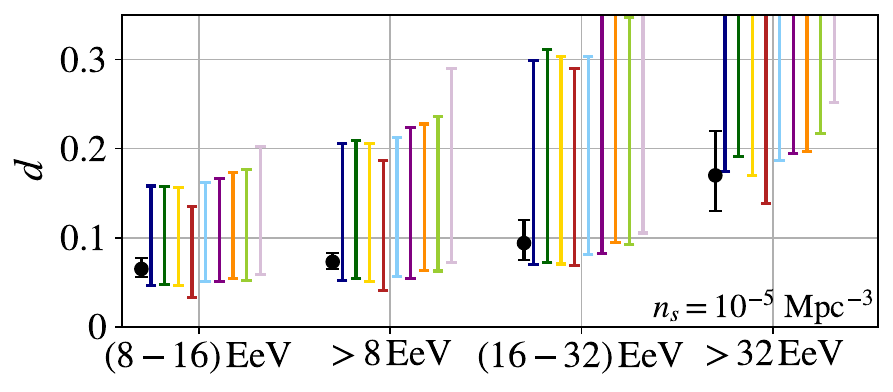} \includegraphics[width=\figwr\textwidth]{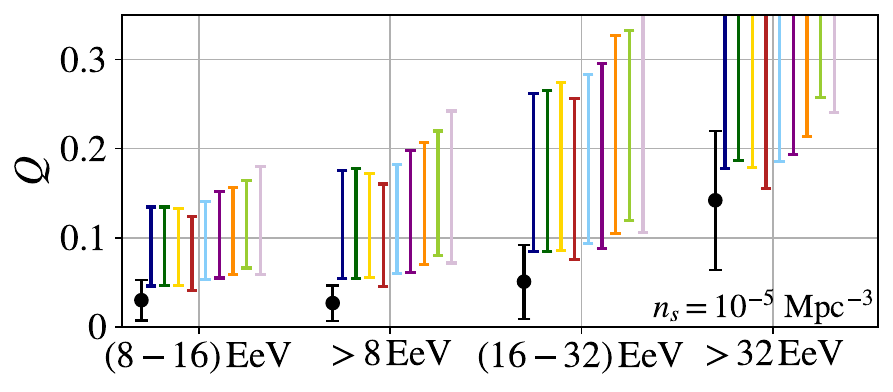}\\
\includegraphics[width=\figwl\textwidth]{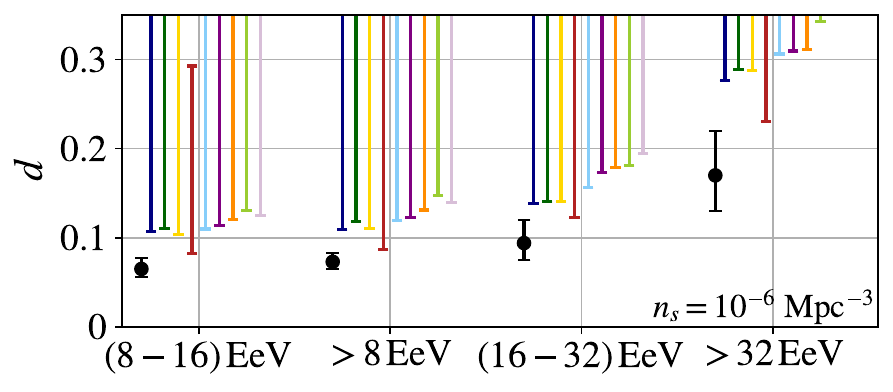} \includegraphics[width=\figwr\textwidth]{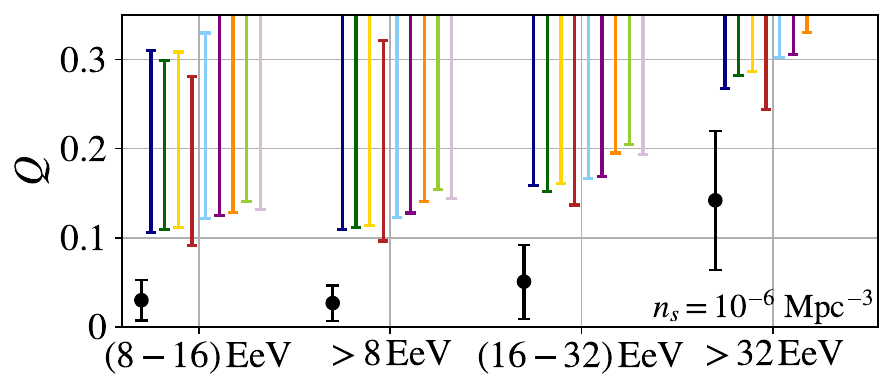}
\caption{\textit{Top row}: the markers show the dipole and quadrupole moments, $d$ and $Q$, in the limit of continuous source number density ($n_s=\infty$) for the various GMF models. The black markers indicate the data $1\sigma$ uncertainty regions for the dipole~\citep{Golup_ADs_2023} and quadrupole~\citep{Caccianiga_ICRC2023}.
\textit{Second - sixth row}: when the density of sources is finite, cosmic variance in the source locations leads to variations of the dipole and quadrupole amplitude, here indicated by the error bar showing the inner $68\%$ of the distribution (for one realization of the model with $l_c=60\,\mathrm{pc}$). For $10^{-5}\,\mathrm{Mpc}^{-3} \lesssim n_s  \lesssim 10^{-3}\,\mathrm{Mpc}^{-3}$, both the \texttt{UF23} dipole and quadrupole moments agree with the data within $1\sigma$ for all models.}
\label{fig:dip_quad_edep}
\end{figure*}

In Fig.~\ref{fig:dip_quad_edep}, we show the dipole and quadrupole amplitudes and their $1\sigma$ uncertainties for various source number densities, for the different GMF models and no extragalactic magnetic field, as a function of energy.

\section{Additional magnification maps} \label{app:more_magnification}
\begin{figure*}[ht]
    \centering
    \subfloat[\texttt{JF12-reg} (compare to~\citep{farrar_sutherland_deflections_2019})]{\includegraphics[trim={0 3.5cm 0 0}, clip, width=5.9cm]{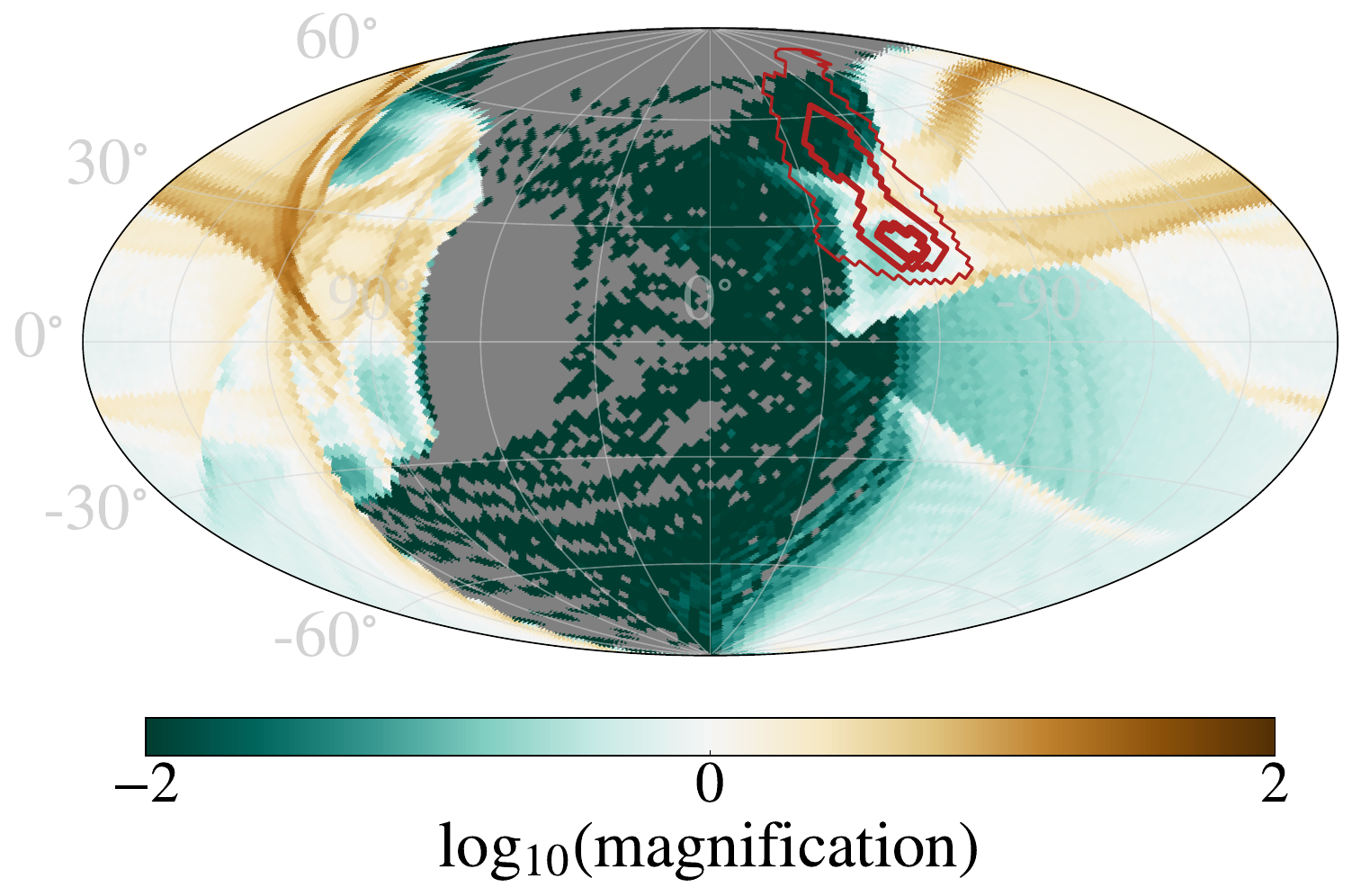}}
    \subfloat[\texttt{JF12-full} $l_c=30\,\mathrm{pc}$ (compare to~\citep{farrar_sutherland_deflections_2019})]{\includegraphics[trim={0 3.5cm 0 0}, clip, width=5.9cm]{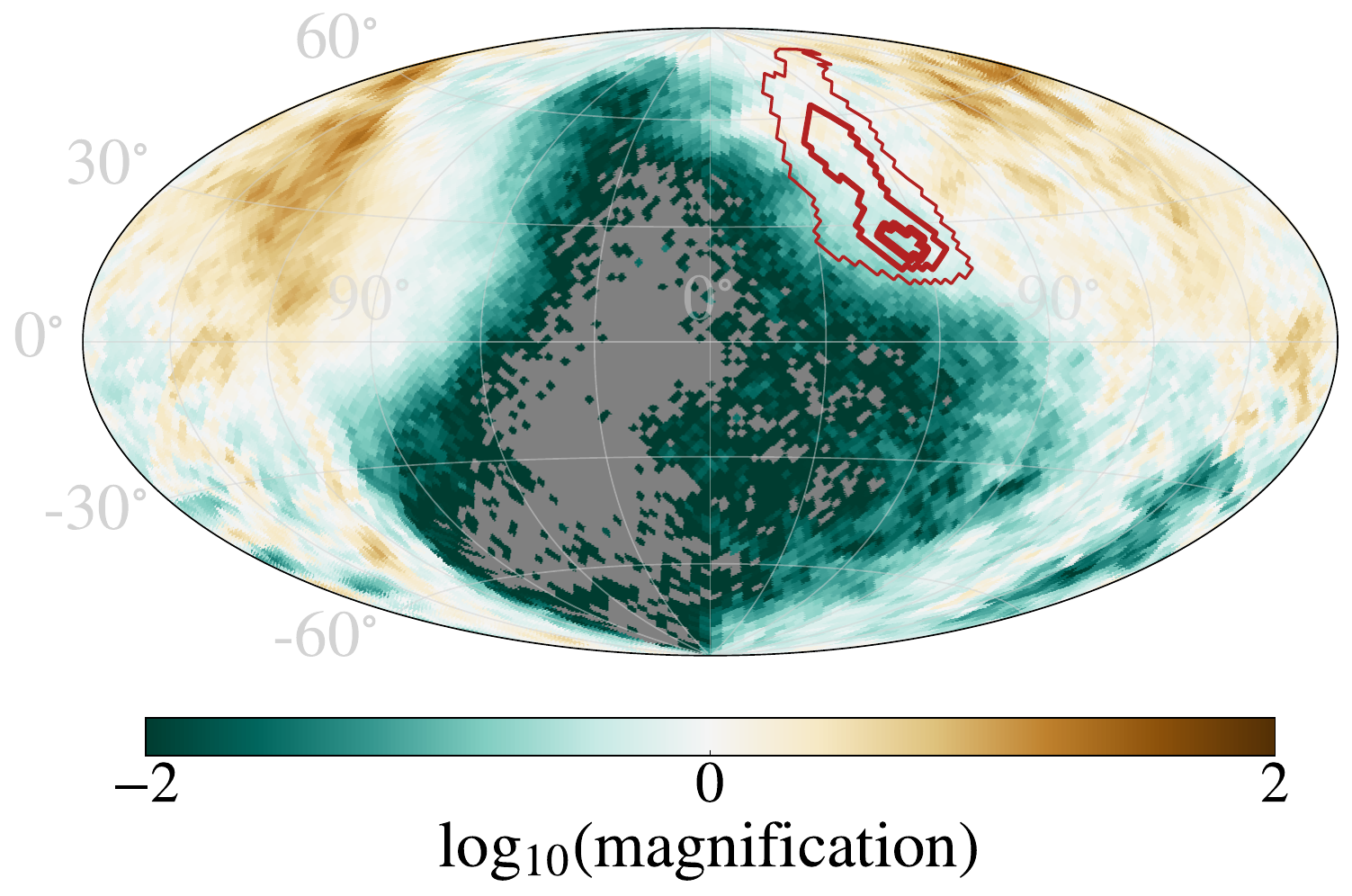}}
    \subfloat[\texttt{JF12} + Planck  $l_c=60\,\mathrm{pc}$]{\includegraphics[trim={0 3.5cm 0 0}, clip, width=5.9cm]{MAGNFICATION_jf12-32-60_5EV.pdf}}
    \\[-0.3ex]

    \subfloat[\texttt{UF23 base} + Planck  $l_c=60\,\mathrm{pc}$]{\includegraphics[trim={0 3.5cm 0 0}, clip, width=5.9cm]{MAGNFICATION_uf23-base-123-lc60-nside32_5EV.pdf}}
    \subfloat[\texttt{UF23 base} + Planck  $l_c=60\,\mathrm{pc}$, \\2nd realization]{\includegraphics[trim={0 3.5cm 0 0}, clip, width=5.9cm]{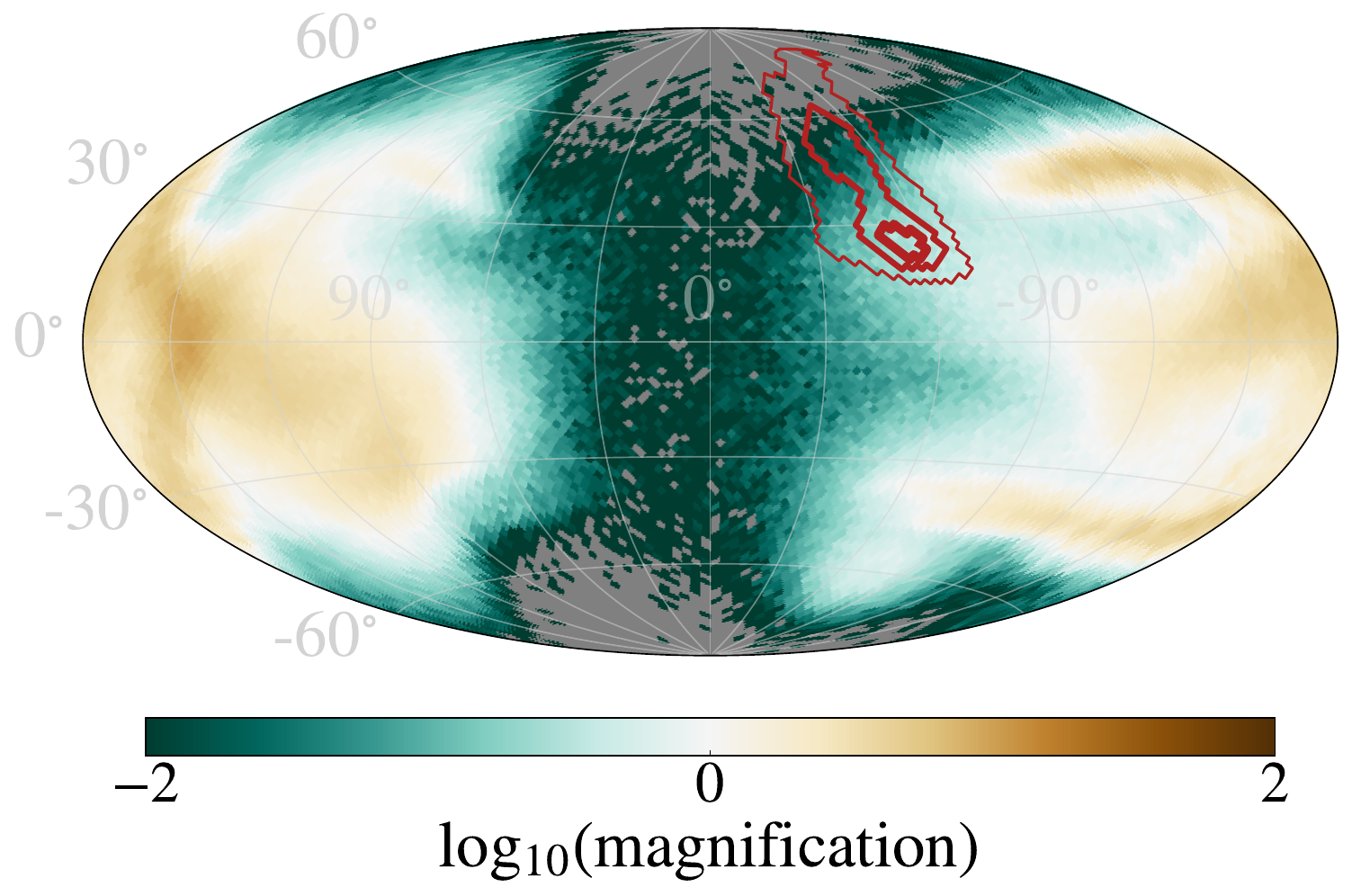}}
    \subfloat[\texttt{UF23 base} + Planck  $l_c=30\,\mathrm{pc}$]{\includegraphics[trim={0 3.5cm 0 0}, clip, width=5.9cm]{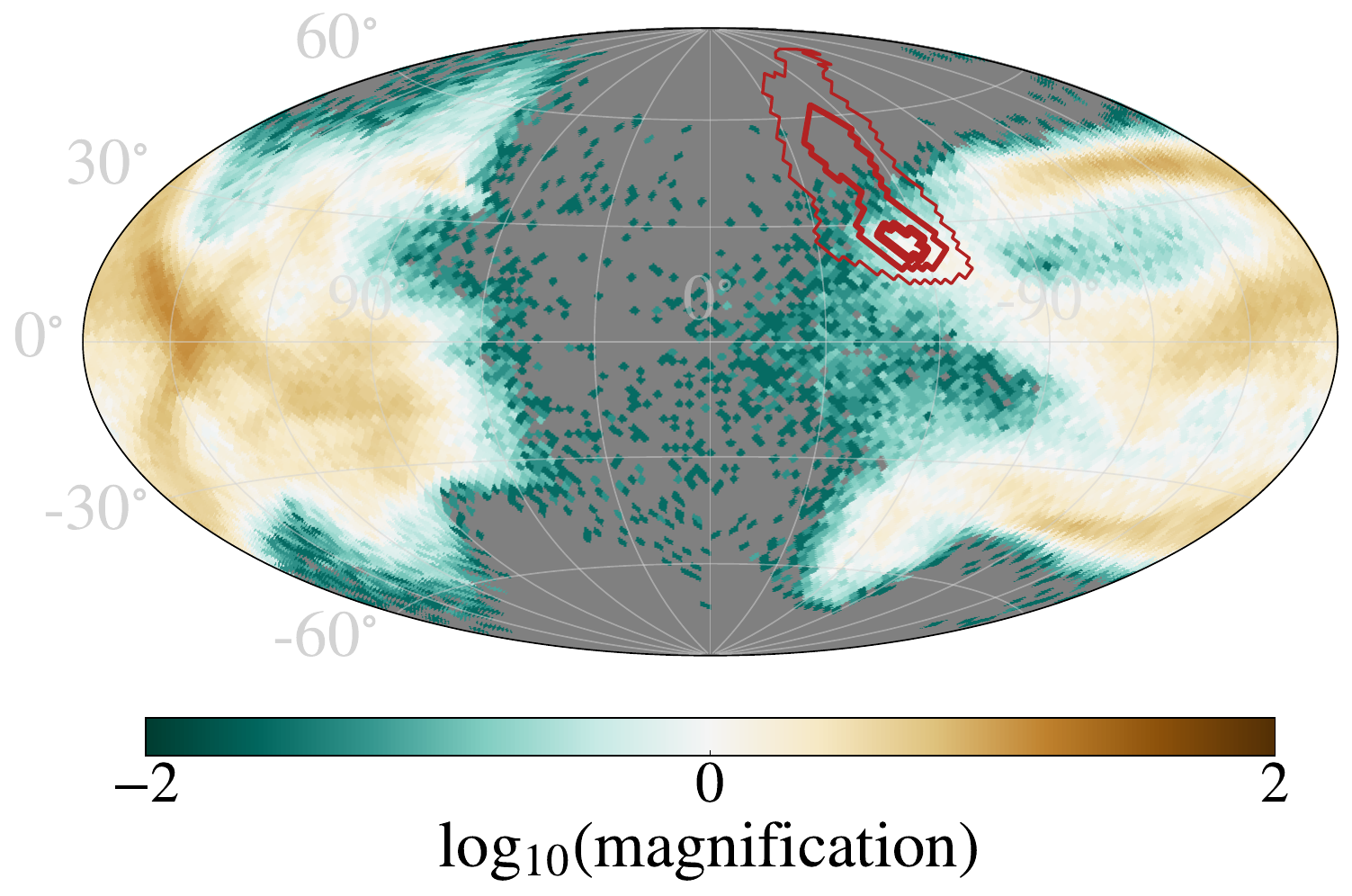}}
    \\[-0.3ex]

    \subfloat[\texttt{UF23 cre10} + Planck  $l_c=60\,\mathrm{pc}$\label{fig:mag_twistx}]{\includegraphics[trim={0 3.5cm 0 0}, clip, width=5.9cm]{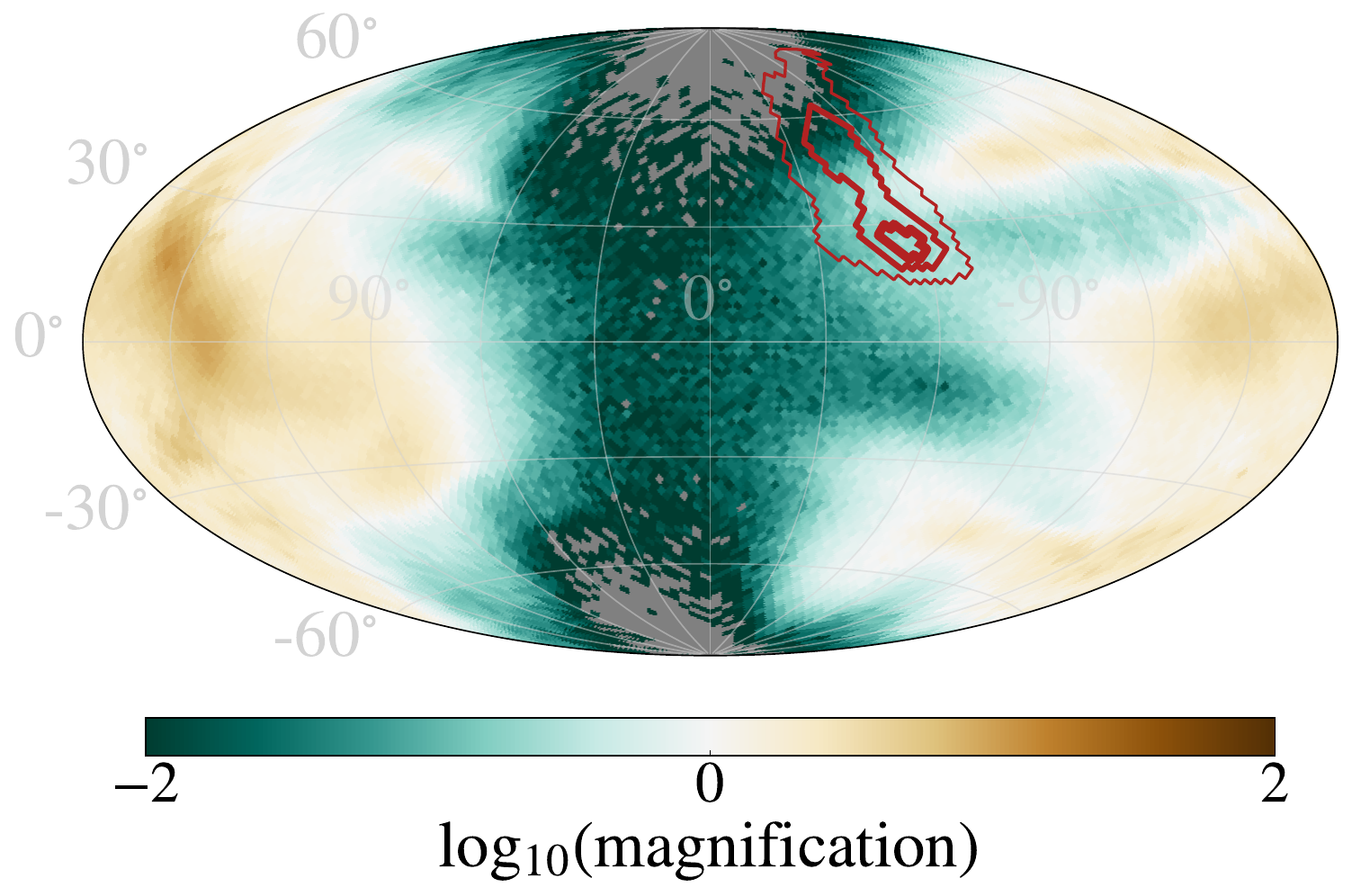}}
    \subfloat[\texttt{UF23 expX} + Planck  $l_c=60\,\mathrm{pc}$]{\includegraphics[trim={0 3.5cm 0 0}, clip, width=5.9cm]{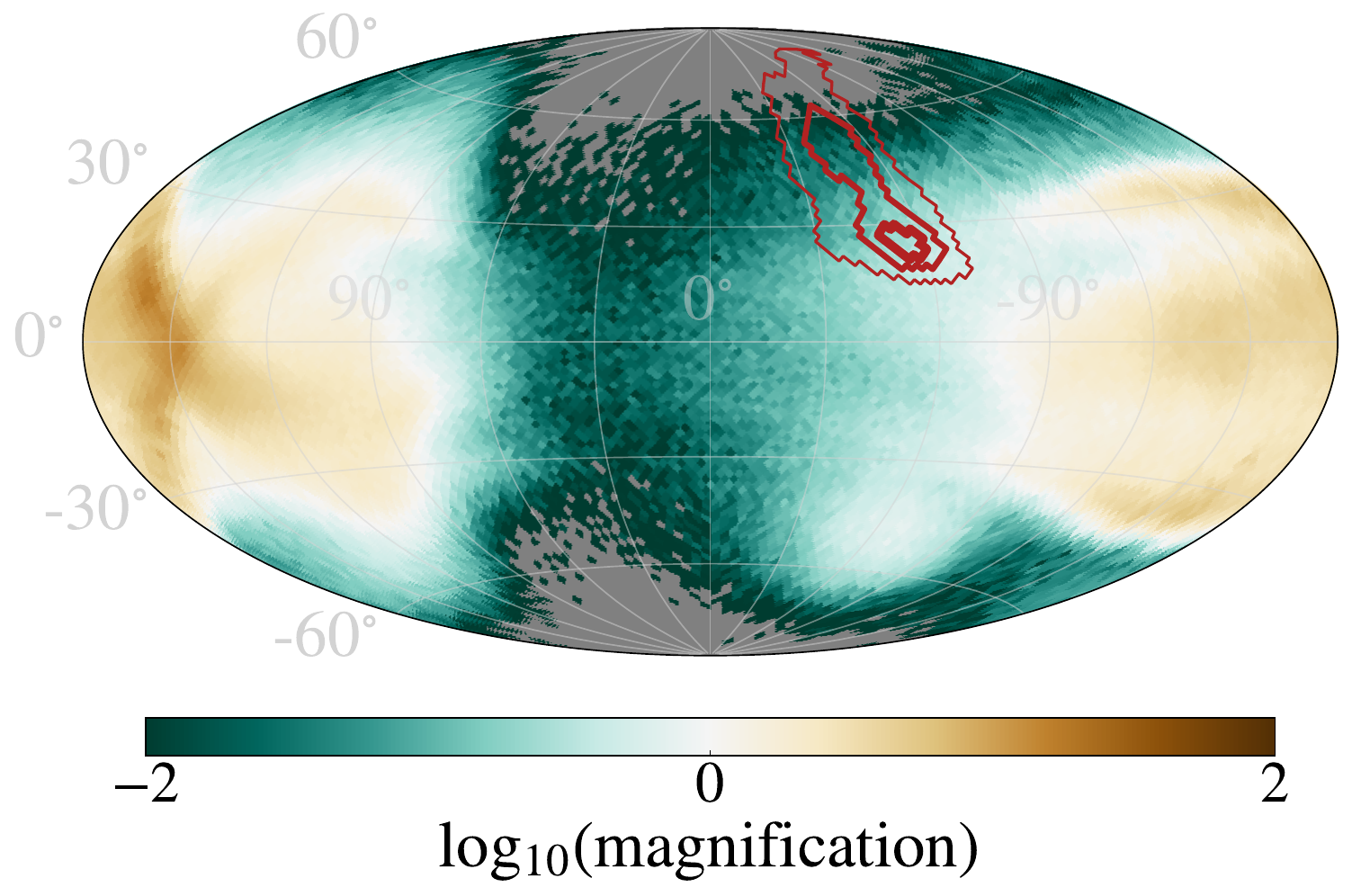}}
    \subfloat[\texttt{UF23 nebCor} + Planck  $l_c=60\,\mathrm{pc}$]{\includegraphics[trim={0 3.5cm 0 0}, clip, width=5.9cm]{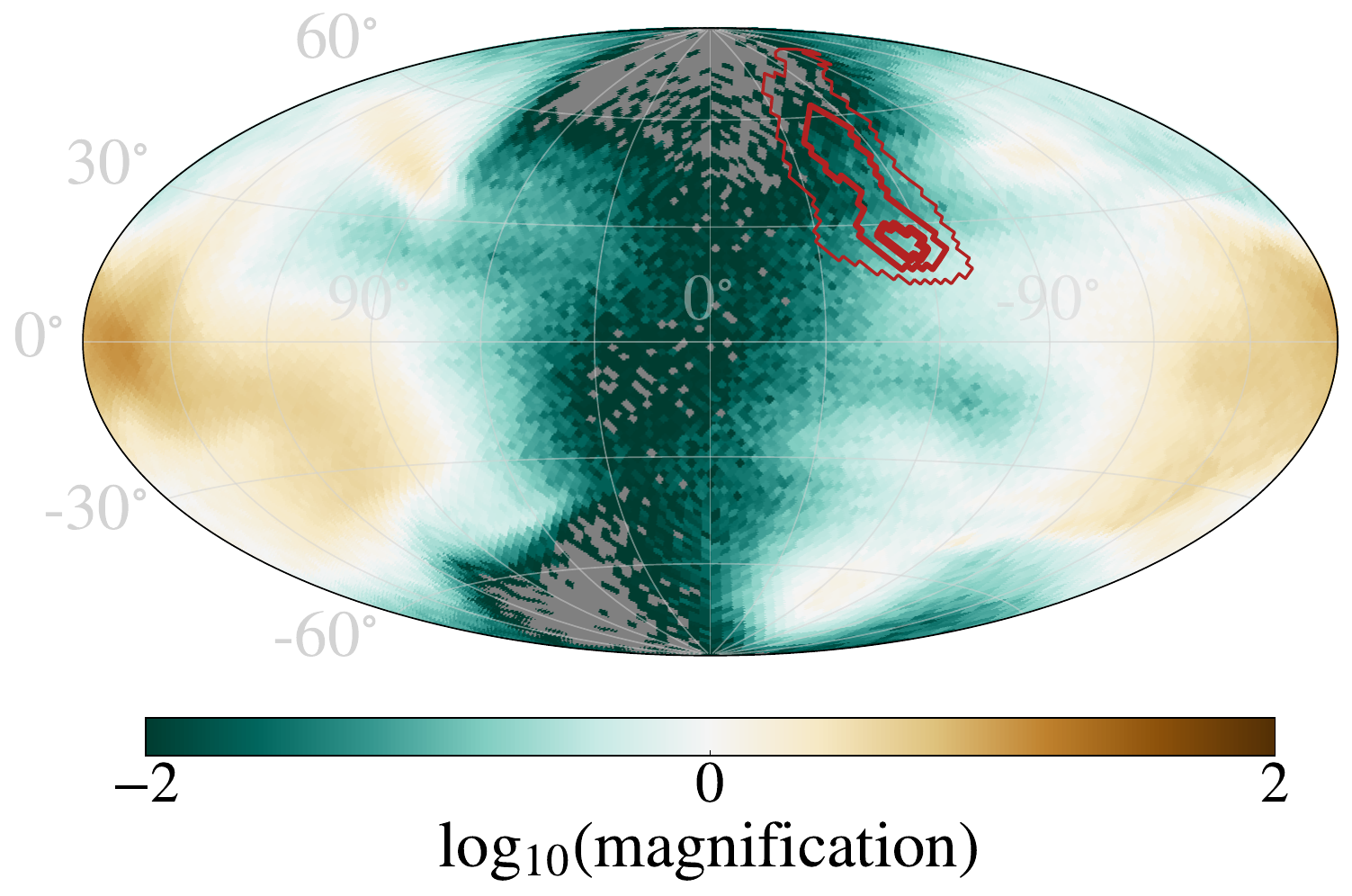}}
    \\[-0.3ex]

    \subfloat[\texttt{UF23 neCL} + Planck  $l_c=60\,\mathrm{pc}$]{\includegraphics[trim={0 3.5cm 0 0}, clip, width=5.9cm]{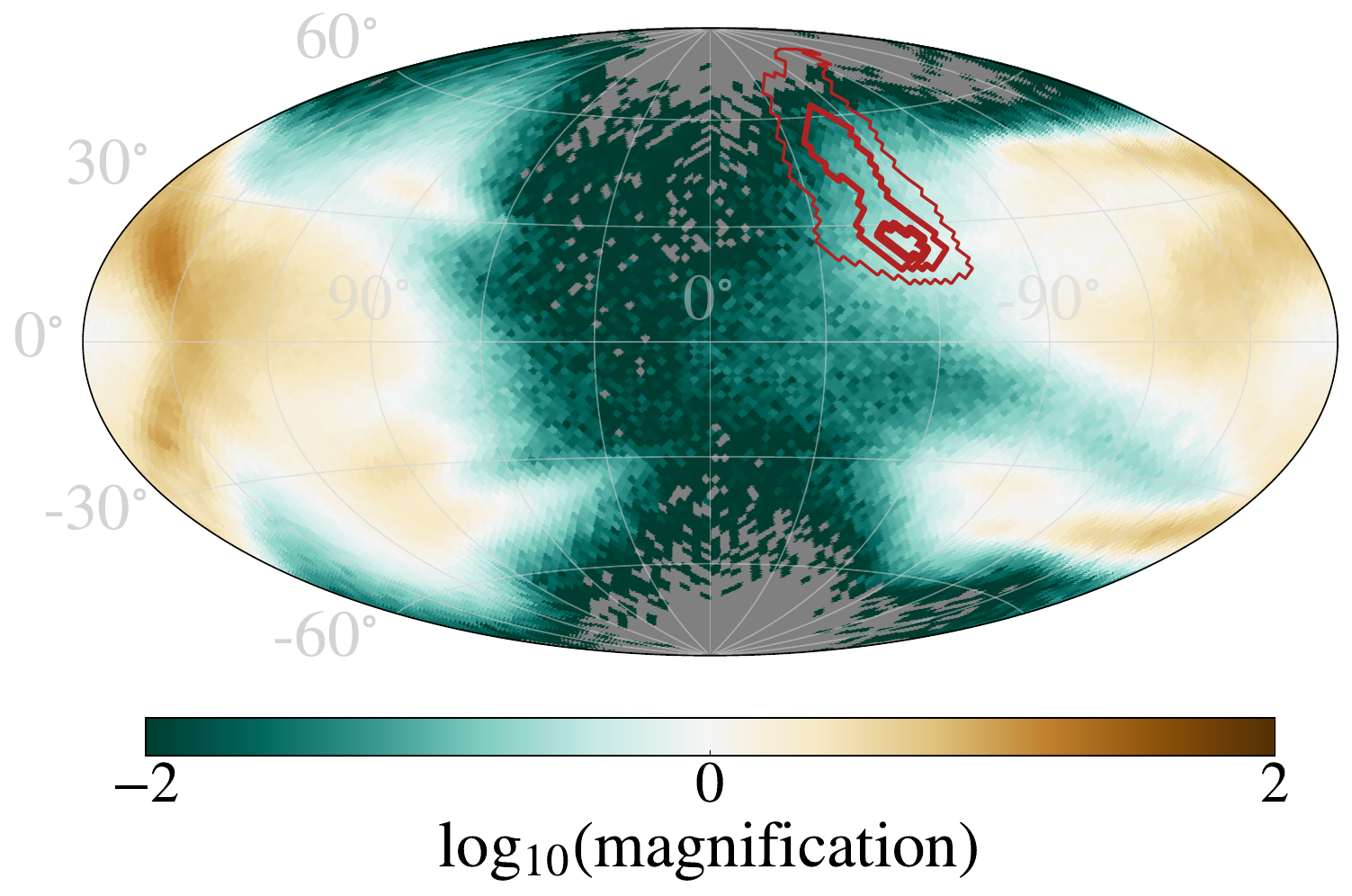}}
    \subfloat[\texttt{UF23 spur} + Planck  $l_c=60\,\mathrm{pc}$]{\includegraphics[trim={0 3.5cm 0 0}, clip, width=5.9cm]{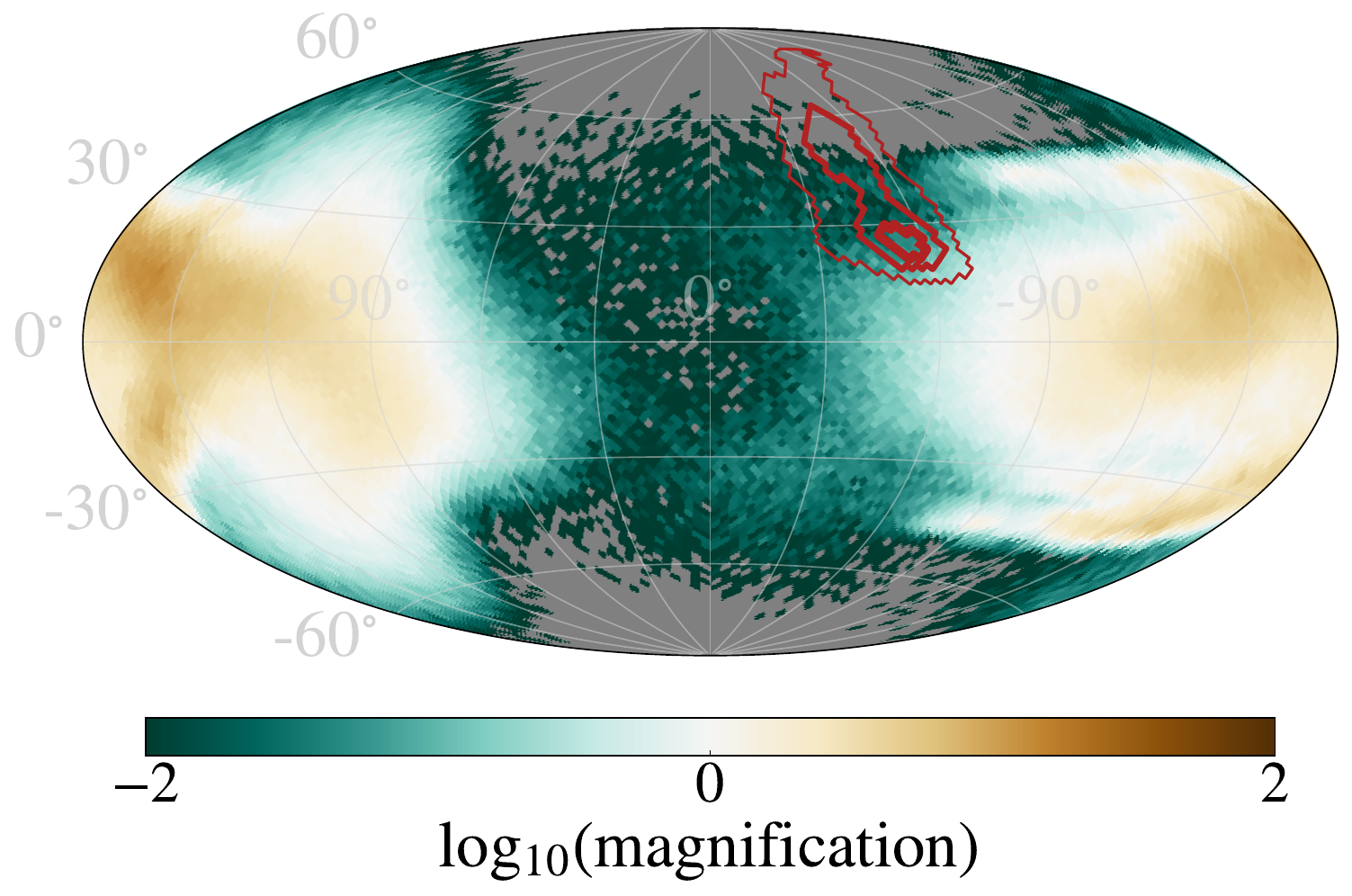}}
    \subfloat[\texttt{UF23 synCG} + Planck  $l_c=60\,\mathrm{pc}$]{\includegraphics[trim={0 3.5cm 0 0}, clip, width=5.9cm]{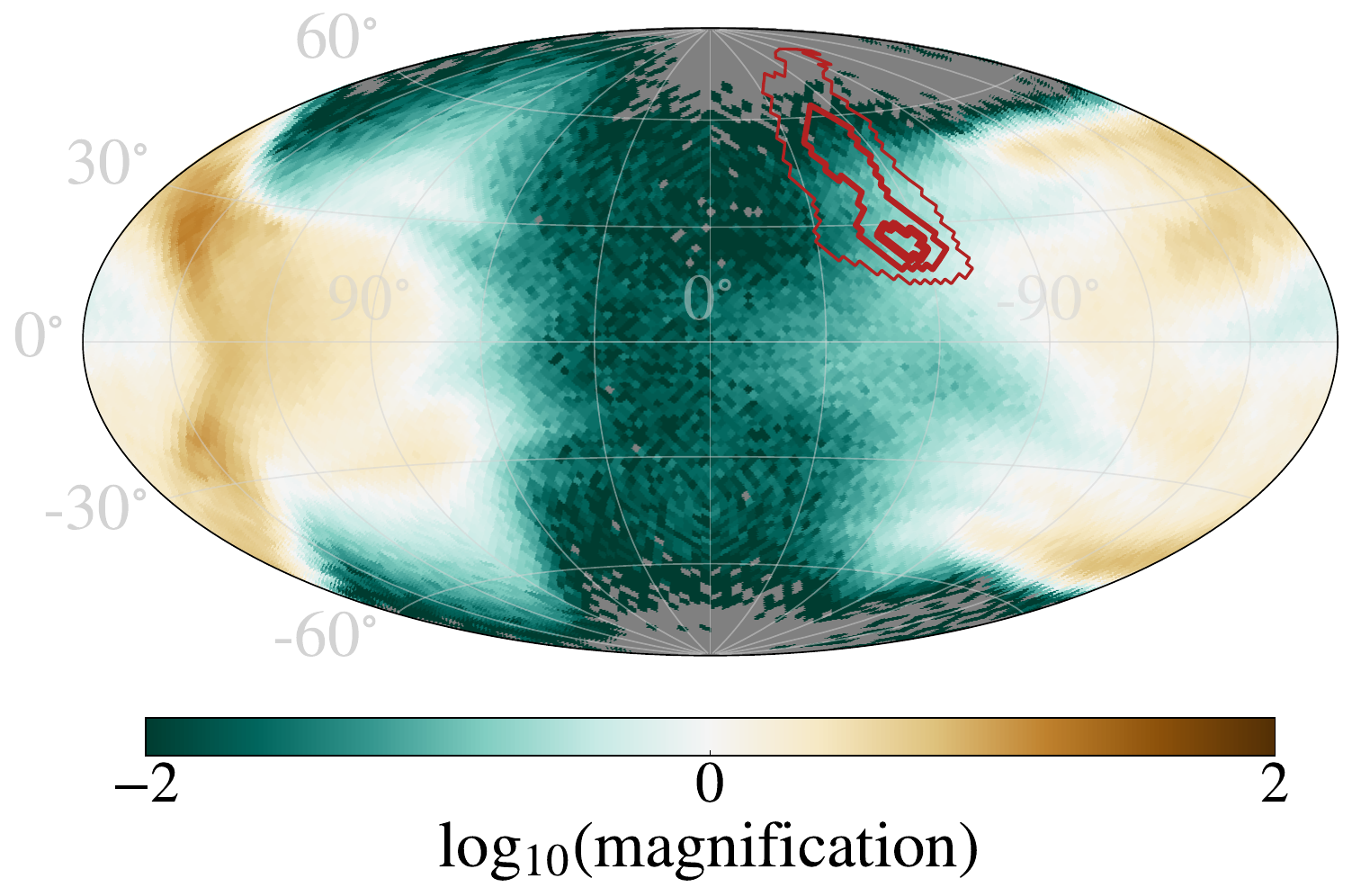}}
    \\[-0.3ex]

    \subfloat[\texttt{UF23 twistX} + Planck  $l_c=60\,\mathrm{pc}$]{\includegraphics[width=5.9cm]{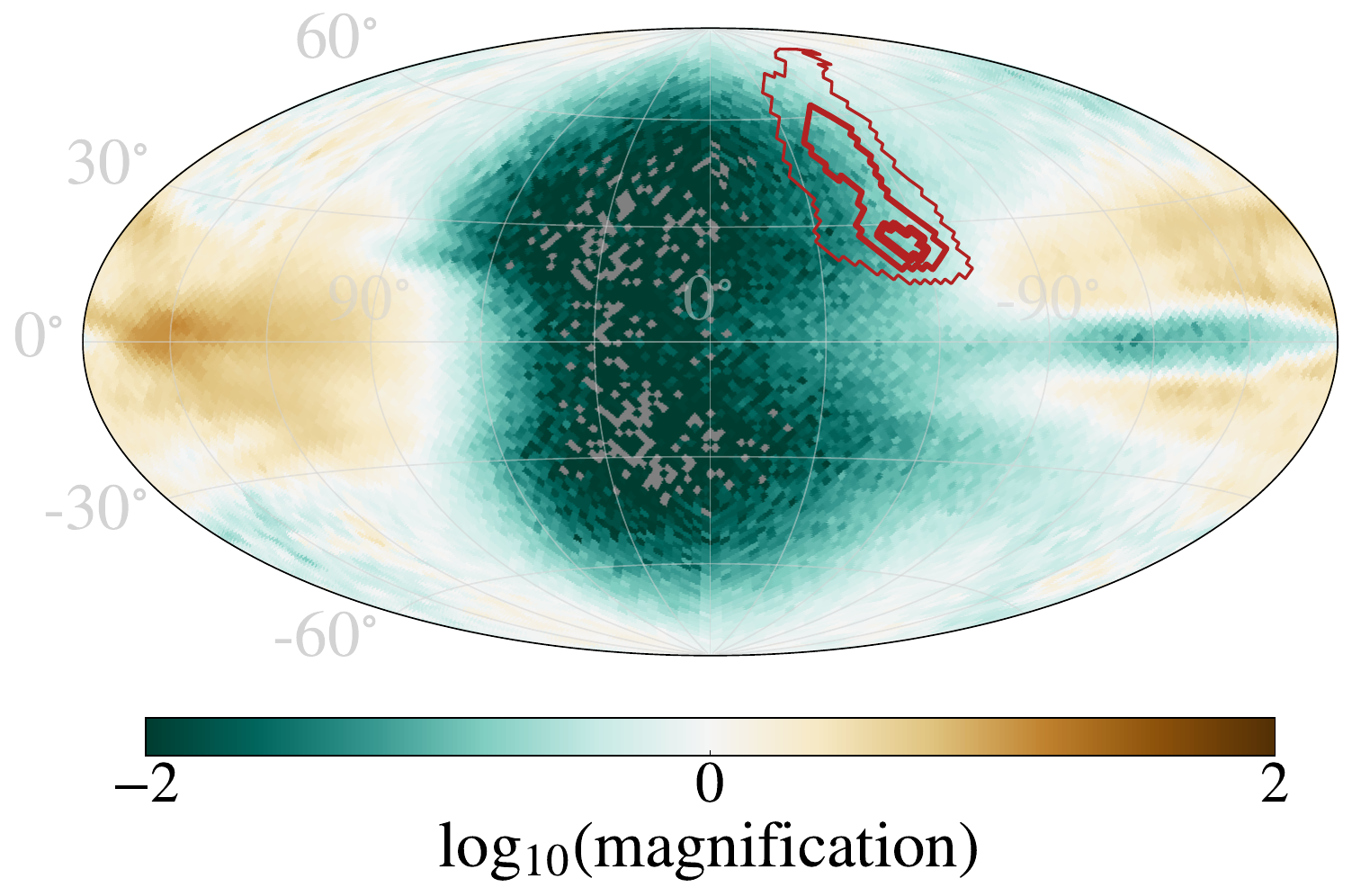}}
    \subfloat[``Illumination" map $I$\\from~\citet{BF23}]{\includegraphics[width=5.9cm]{best_fit_dipole__I_map_9_-1.pdf}}

    \caption{a)-m): Magnification maps for rigidity $\mathcal{R}=5\,\mathrm{EV}$ (as in Fig.~\ref{fig:magnification_maps}) for various GMF models including also variations of the random field.
    Contours indicating the extragalactic directions with large flux predicted by the LSS model (panel (n)) are shown in red.
    n): The $E>8\,\mathrm{EeV}$ illumination map calculated from the LSS model~\citep{BF23}, showing the flux at the edge of the Galaxy.}
    \label{fig:magnification_maps_more}
\end{figure*}

In Fig.~\ref{fig:magnification_maps_more}, the magnification maps for all \texttt{UF23} models with $l_c=60\,\mathrm{pc}$, as well as for the \texttt{JF12} models with different random field models are shown.
To determine how much the uncertainty on the random field influences the magnification map, we show in the second row the magnification maps for the \texttt{base} model with the three tested random fields (two realizations with $l_c=60\,\mathrm{pc}$ and one with $l_c=30\,\mathrm{pc}$). The particular realization of the random field makes almost no difference to the magnification map, while the coherence length can have a visible impact. Especially the exact form and size of the demagnified region are sensitive to the coherence length.

\begin{figure*}[ht]
  \centering
\def\figw{0.32}
\includegraphics[width=\figw\linewidth]{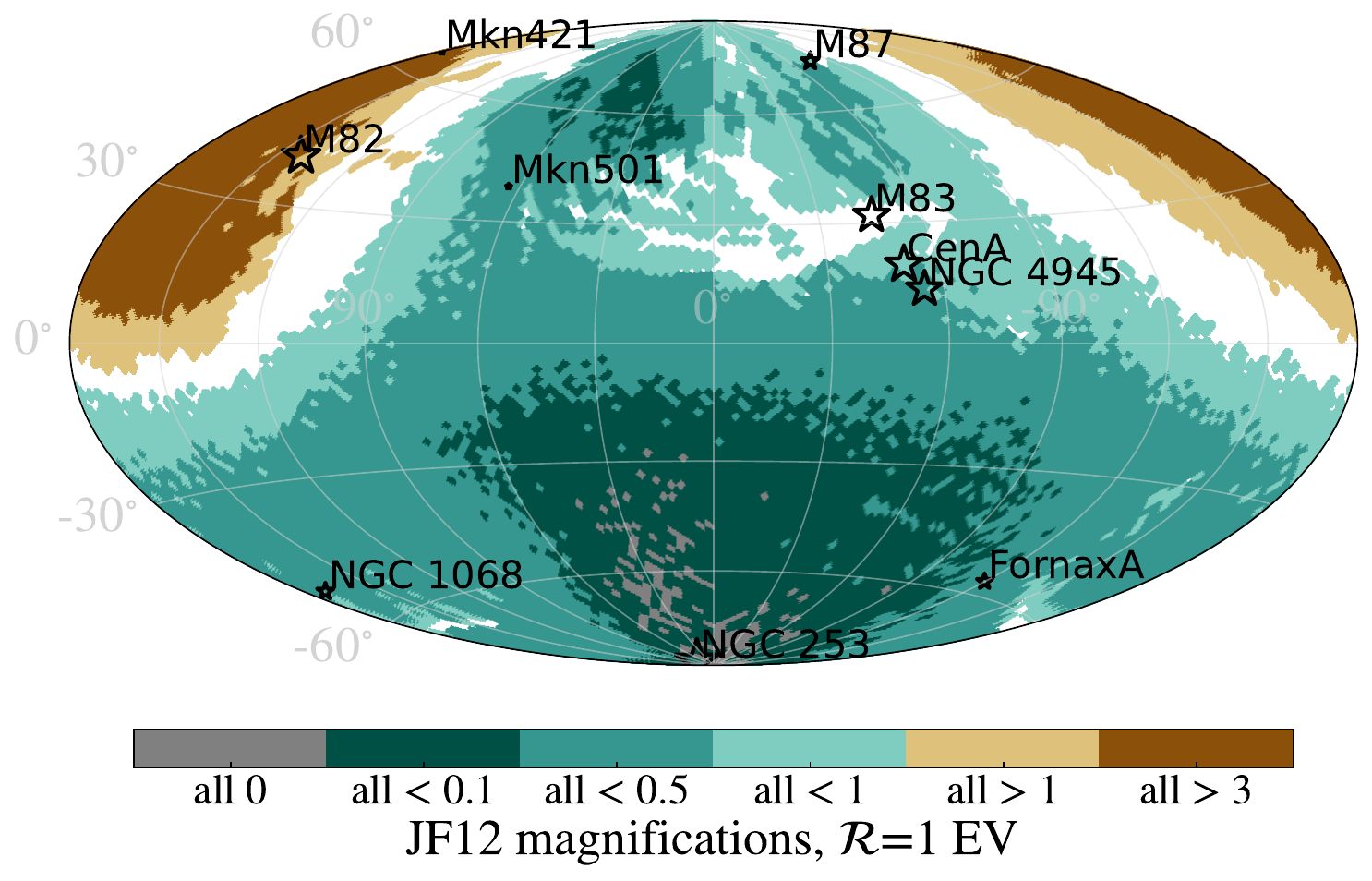}
\includegraphics[width=\figw\linewidth]{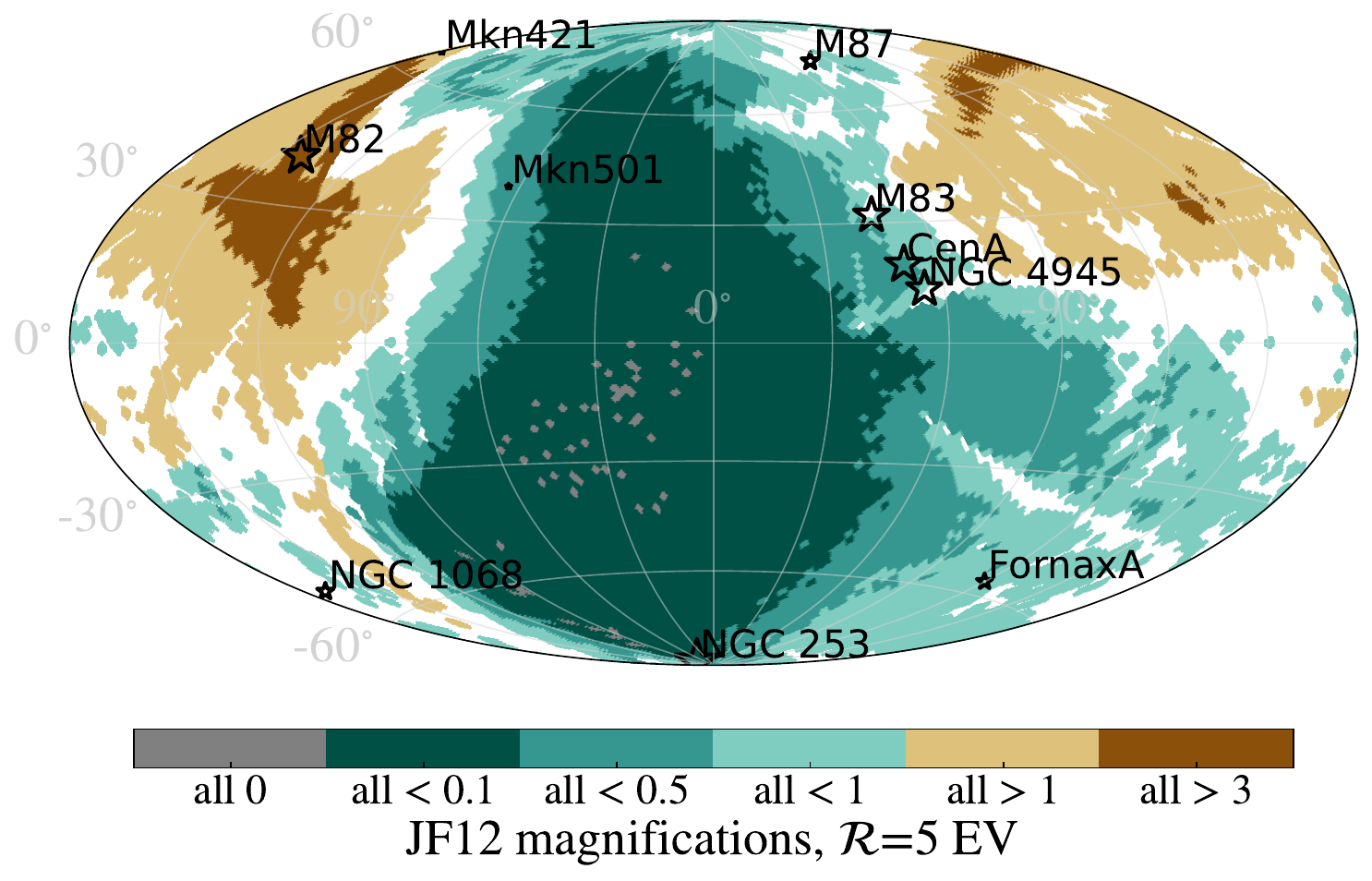}
\includegraphics[width=\figw\linewidth]{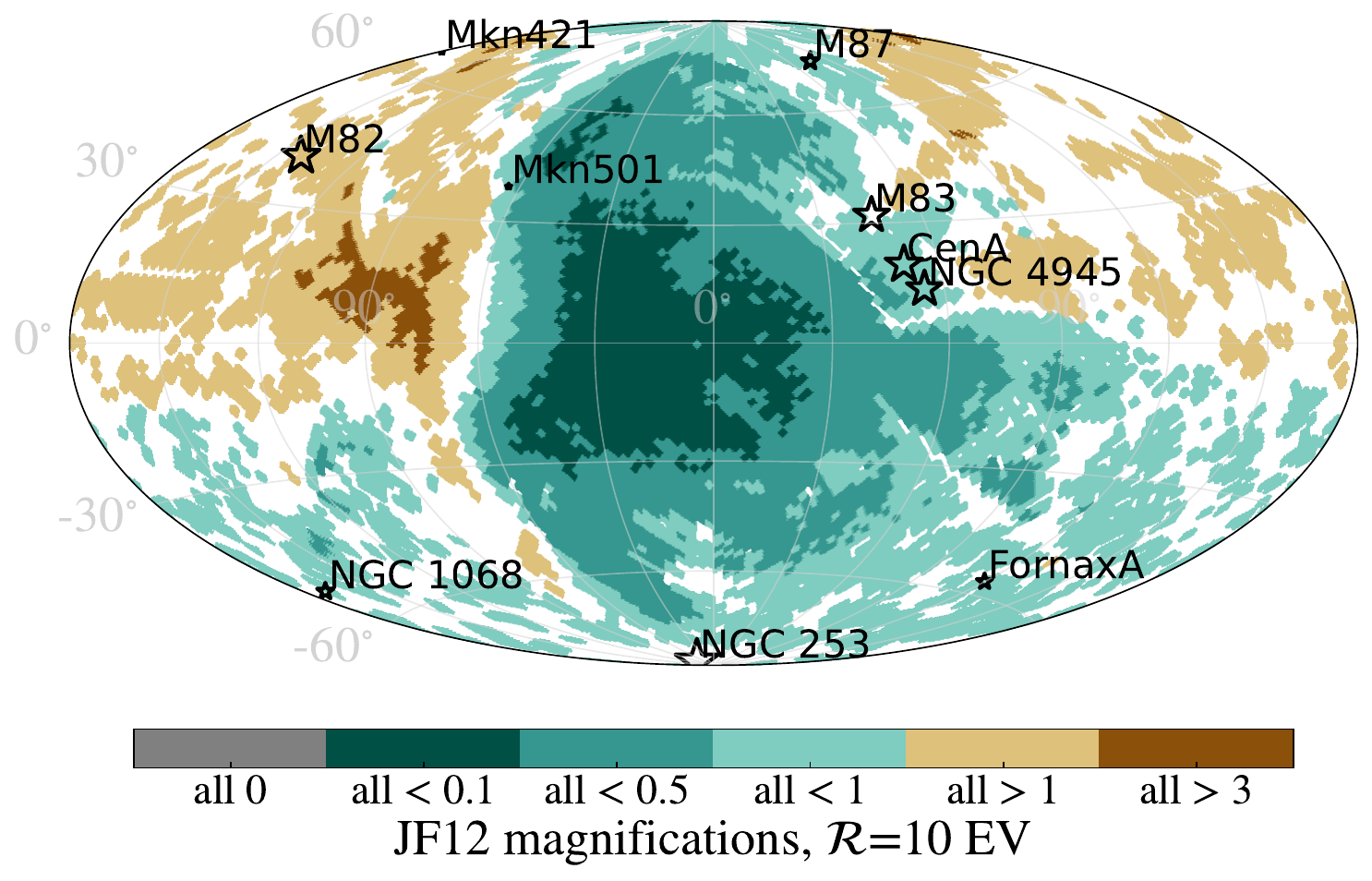}
\caption{Combined magnification maps (as in Fig.~\ref{fig:magnification_combined}) for the \texttt{JF12} model with the different random fields used in this work (\texttt{JF12-regular}, \texttt{JF12-full} with $l_c=30\,\mathrm{pc}$ and \texttt{JF12-Planck} with $l_c=60\,\mathrm{pc}$, the latter with two variations), illustrating the sensitivity of the magnification to the random field and -- by comparing to Fig.~\ref{fig:magnification_combined} -- the general differences between the \texttt{JF12} and \texttt{UF23} coherent models as a function of rigidity.
The color bar displays the magnification range in directions where all combinations of \texttt{JF12} and random models  agree; for the white area there is no consensus among the models. The directions of source candidates are indicated by stars and the marker size is proportional to $1/\mathrm{distance}$.}
\label{fig:magnification_combined_jf12}
\end{figure*}

Fig.~\ref{fig:magnification_combined_jf12} displays the combined magnification maps (see sec.~\ref{sec:magnification} for explanation) for all combinations of \texttt{JF12} and random field models used 
in this work. It is visible that the areas of magnification and demagnification are distinctly different as for the \texttt{UF23} models. Many of the depicted source candidates like Mkn501 and NGC253 again lie in the demagnification area for rigidities $\mathcal{R}\leq5\,\mathrm{EV}$ -- but, sources in the equatorial North like Mkn421 or M82 are significantly magnified with \texttt{JF12} in contrast to \texttt{UF23}.

\section{Idealized extragalactic dipole} \label{app:dipoleI}
\begin{figure*}[ht]
\includegraphics[width=0.6\textwidth]{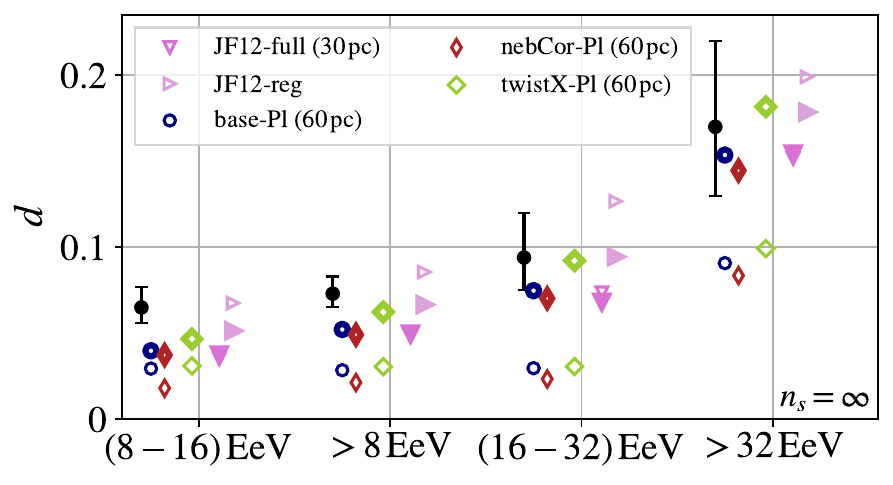}
\caption{Dipole amplitude dependence on the extragalactic flux: the thin markers indicate the predicted dipole amplitude for the LSS model~\citep{BF23} for selected GMF models. The respective thicker markers of the same form show the dipole amplitude prediction when the LSS illumination map is replaced by a smooth dipole with the same magnitude and direction -- demonstrating the sensitivity of the observed dipole amplitude to the inhomogeneities in the illumination map. The black markers represent the measured dipole amplitude and its $1\sigma$ uncertainty~\citep{Golup_ADs_2023}.}
\label{fig:dipoleI_amp}
\end{figure*}

In Fig.~\ref{fig:dipoleI_amp} we show the dipole amplitude predicted by the LSS model, compared to a model where the flux at the edge of the galaxy (the ``illumination map", Fig.~\ref{fig:illum}) is replaced by an idealized dipole with the same amplitude and direction as for the LSS model. The amplitude of the LSS dipole at the edge of the galaxy is $6.2\%$ for the $E=(8-16)\,\mathrm{EeV}$ energy bin, $8.1\%$ for $E>8\,\mathrm{EeV}$, $11.9\%$ for $E=(16-32)\,\mathrm{EeV}$, and $21.5\%$ for $E>32\,\mathrm{EeV}$~\citep{BF23}.
It is visible that the dipole amplitude is highly sensitive to the inhomogeneities in the extragalactic flux. For all shown \texttt{UF23} models, the amplitude of the LSS model is significantly smaller (around a factor 2) than for the model with idealized extragalactic dipole, while this relation is the other way around for the \texttt{JF12-reg} model. This is due to the intricate relation between the illumination and the magnification of the GMF, which differs significantly between the \texttt{UF23} and \texttt{JF12} models as explained in the main text.

\begin{figure*}[ht]
\subfloat[energy $>8\,\mathrm{EeV}$]{\includegraphics[height=4cm]{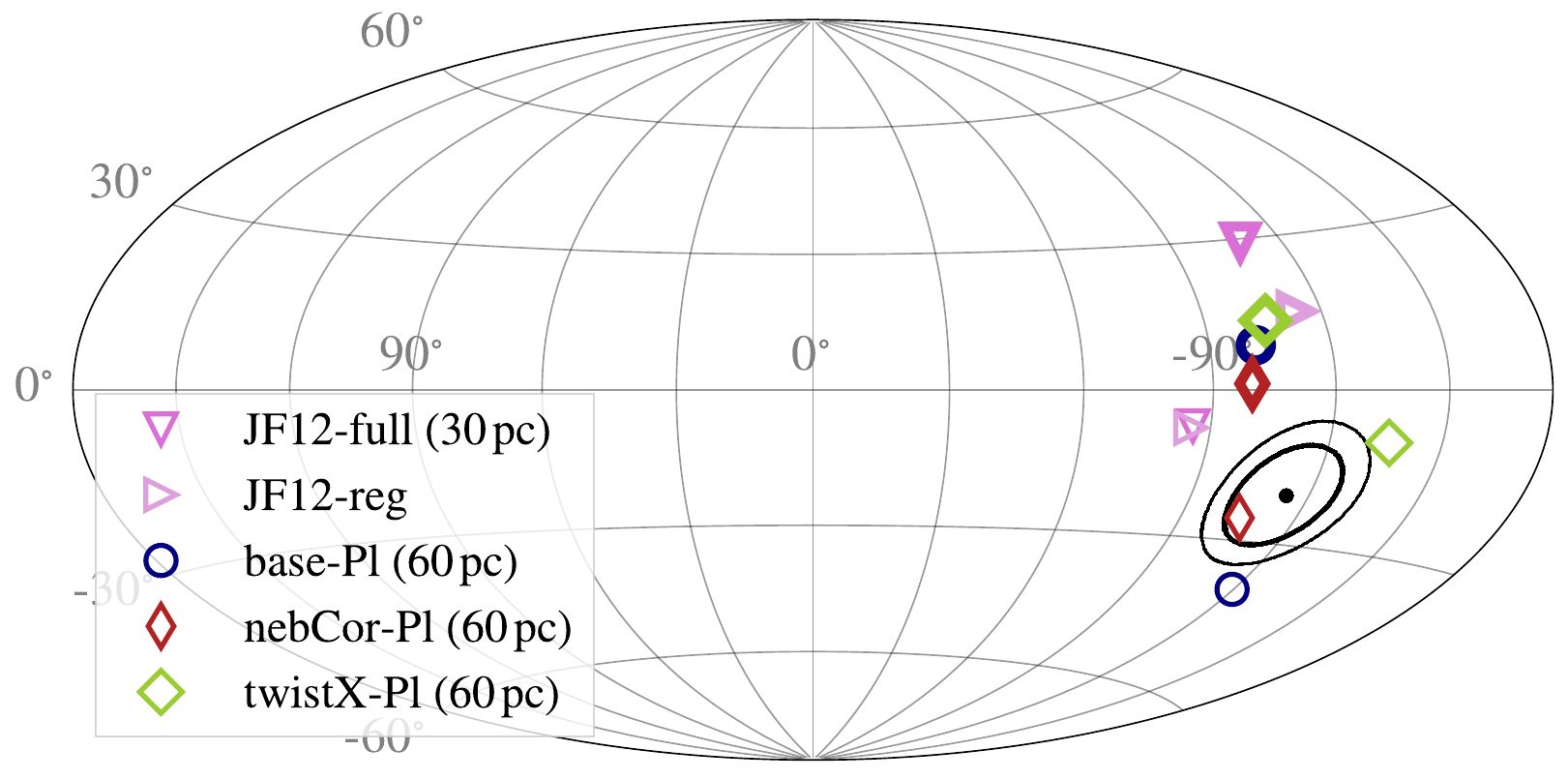}}
\subfloat[$(8-16)\,\mathrm{EeV}$]{\includegraphics[trim={16cm 0 0 0}, clip, height=4cm]{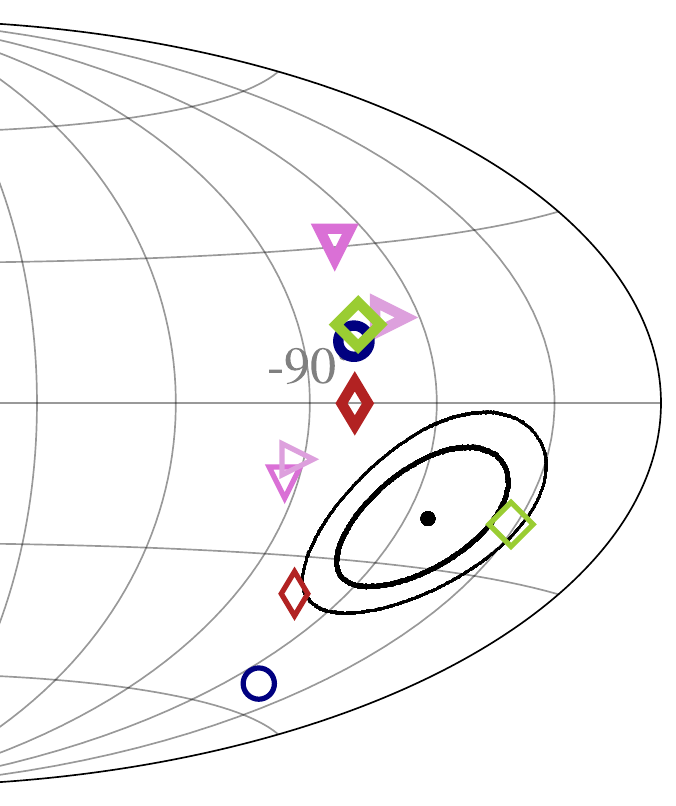}}
\subfloat[$(16-32)\,\mathrm{EeV}$]{\includegraphics[trim={16cm 0 0 0}, clip, height=4cm]{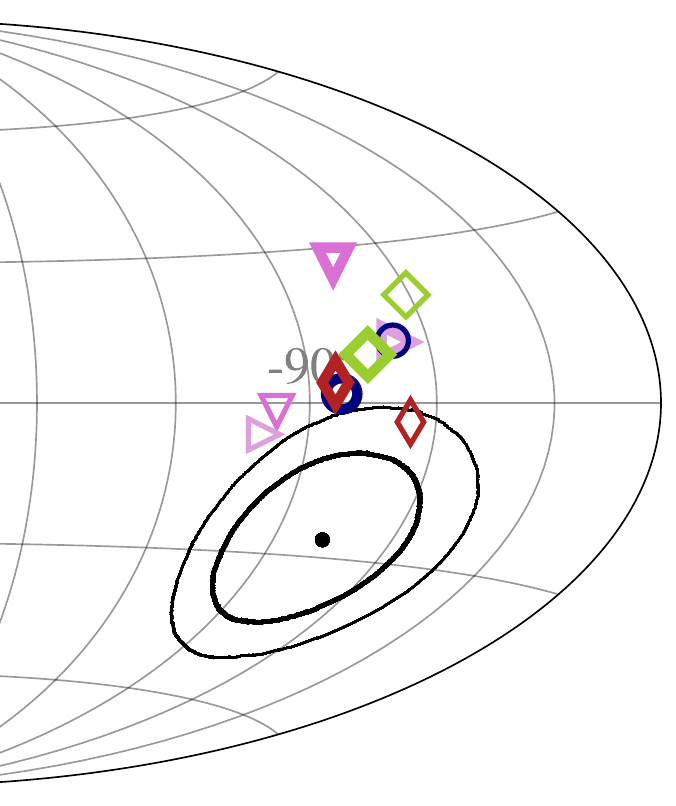}}
\subfloat[$>32\,\mathrm{EeV}$]{\includegraphics[trim={16cm 0 0 0}, clip, height=4cm]{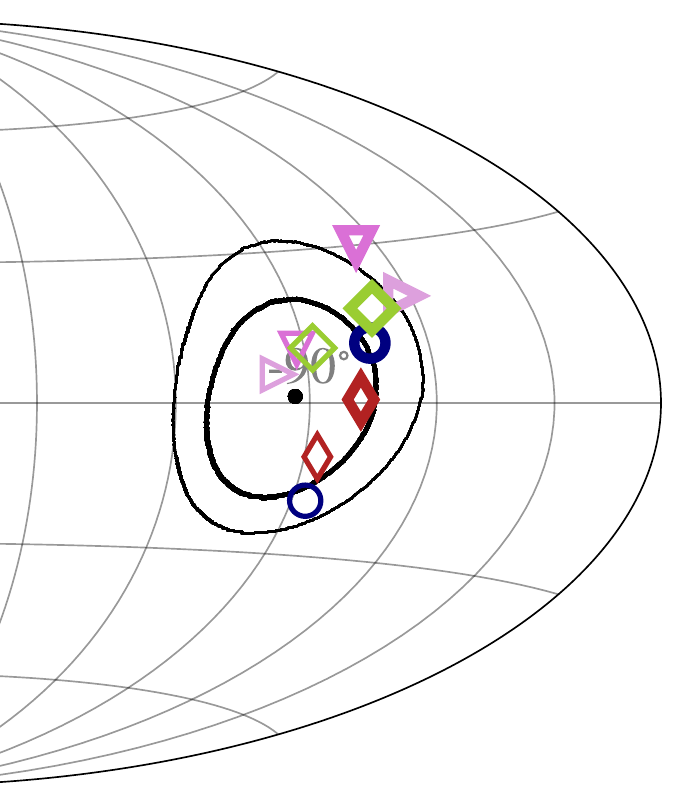}}
\caption{Dipole direction dependence on the extragalactic flux, for different energy thresholds in Galactic coordinates:
the thin markers indicate the predicted dipole directions for the LSS model~\citep{BF23} for selected GMF models. The respective thicker  markers of the same form show the dipole direction prediction when the LSS illumination map is replaced by a smooth dipole with the same magnitude and direction -- demonstrating the sensitivity of the observed dipole direction to the inhomogeneities in the illumination map.
The black contours represent the $1\sigma$ and $2\sigma$ uncertainty domains of the measured dipole~\citep{Golup_ADs_2023}.}
\label{fig:dipoleI}
\end{figure*}

In Fig.~\ref{fig:dipoleI}, we show the dipole directions calculated when replacing the flux at the edge of the galaxy predicted by the LSS model by the idealized dipole. The direction of the dipole calculated at Earth differs substantially -- by $\mathcal{O}(20^\circ \,{\rm to}\,60^\circ)$ -- between this simplification and the realistic model where the sources follow the LSS. Also, the direction predicted using the idealized dipole is systematically displaced towards the North, especially for lower energies, and moves significantly less with the energy than for the LSS model. Comparing the prediction for the \texttt{base} model with idealized dipole to the uncertainty contour from cosmic variance for $n_s=10^{-3}\,\mathrm{Mpc}^{-3}$ (Fig.~\ref{fig:dipole_direcs}), it is visible that the predicted dipole direction of the idealized dipole model is even outside that sizable uncertainty for lower energies.

This test demonstrates that simply replacing an extragalactic source catalog by its dipole component (the ``2MRS dipole" which is often used in the literature, e.g.~\citet{Auger_dipole_2017_Science, Auger_dipole_2018, Bray_2018, Bakalova_2023}) and ignoring higher multipoles of the distribution, can only give a rough idea of the deflection direction and expected dipole amplitude, but should not be expected to produce accurate predictions of the expected anisotropy at Earth.


\bibliography{bibliography}{}

\begin{thebibliography}{}
\expandafter\ifx\csname natexlab\endcsname\relax\def\natexlab#1{#1}\fi
\providecommand{\url}[1]{\href{#1}{#1}}
\providecommand{\dodoi}[1]{doi:~\href{http://doi.org/#1}{\nolinkurl{#1}}}
\providecommand{\doeprint}[1]{\href{http://ascl.net/#1}{\nolinkurl{http://ascl.net/#1}}}
\providecommand{\doarXiv}[1]{\href{https://arxiv.org/abs/#1}{\nolinkurl{https://arxiv.org/abs/#1}}}

\bibitem[{{A. Yushkov for the Pierre Auger Collaboration}(2019)}]{a_yushkov_for_the_pierre_auger_collaboration_mass_2019}
{A. Yushkov for the Pierre Auger Collaboration}. 2019, in {PoS}({ICRC2019}), Vol. 358, 482, \dodoi{10.22323/1.358.0482}

\bibitem[{Abbott {et~al.}(2023)}]{LVK_binaryMergerRate23}
Abbott, R., {et~al.} 2023, Phys. Rev. X, 13, 041039, \dodoi{10.1103/PhysRevX.13.041039}

\bibitem[{{Achterberg} {et~al.}(1999){Achterberg}, {Gallant}, {Norman}, \& {Melrose}}]{Achterberg:1999vr}
{Achterberg}, A., {Gallant}, Y.~A., {Norman}, C.~A., \& {Melrose}, D.~B. 1999.
\newblock \doarXiv{astro-ph/9907060}

\bibitem[{Ajello {et~al.}(2013)}]{Ajello_2013}
Ajello, M., {et~al.} 2013, ApJ, 780, 73, \dodoi{10.1088/0004-637x/780/1/73}

\bibitem[{Allard {et~al.}(2022)Allard, Aublin, Baret, \& Parizot}]{allard_what_2022}
Allard, D., Aublin, J., Baret, B., \& Parizot, E. 2022, A\&A, 664, A120, \dodoi{10.1051/0004-6361/202142491}

\bibitem[{Andreoni {et~al.}(2022)}]{Andreoni_2022}
Andreoni, I., {et~al.} 2022, Nature, 612, 430, \dodoi{10.1038/s41586-022-05465-8}

\bibitem[{Bakalová {et~al.}(2023)Bakalová, Vícha, \& Trávníček}]{Bakalova_2023}
Bakalová, A., Vícha, J., \& Trávníček, P. 2023, JCAP, 2023, 016, \dodoi{10.1088/1475-7516/2023/12/016}

\bibitem[{Best \& Heckman(2012)}]{Best_2012}
Best, P.~N., \& Heckman, T.~M. 2012, MNRAS, 421, 1569, \dodoi{10.1111/j.1365-2966.2012.20414.x}

\bibitem[{Bister \& Farrar(2024)}]{BF23}
Bister, T., \& Farrar, G.~R. 2024, ApJ, 966, 71, \dodoi{10.3847/1538-4357/ad2f3f}

\bibitem[{Bray \& Scaife(2018)}]{Bray_2018}
Bray, J.~D., \& Scaife, A. M.~M. 2018, ApJ, 861, 3, \dodoi{10.3847/1538-4357/aac777}

\bibitem[{{Caccianiga, L. for the Pierre Auger and Telescope Array Collaborations}(2023)}]{Caccianiga_ICRC2023}
{Caccianiga, L. for the Pierre Auger and Telescope Array Collaborations}. 2023, in PoS(ICRC2023), Vol. 444, 521, \dodoi{10.22323/1.444.0521}

\bibitem[{Condorelli {et~al.}(2023)Condorelli, Biteau, \& Adam}]{Condorelli_2023}
Condorelli, A., Biteau, J., \& Adam, R. 2023, ApJ, 957, 80, \dodoi{10.3847/1538-4357/acfeef}

\bibitem[{Conselice {et~al.}(2016)Conselice, Wilkinson, Duncan, \& Mortlock}]{Conselice_2016}
Conselice, C.~J., Wilkinson, A., Duncan, K., \& Mortlock, A. 2016, ApJ, 830, 83, \dodoi{10.3847/0004-637x/830/2/83}

\bibitem[{di~Matteo \& Tinyakov(2018)}]{di_Matteo_2018}
di~Matteo, A., \& Tinyakov, P. 2018, MNRAS, 476, 715–723, \dodoi{10.1093/mnras/sty277}

\bibitem[{Ding {et~al.}(2021)Ding, Globus, \& Farrar}]{ding_imprint_2021}
Ding, C., Globus, N., \& Farrar, G.~R. 2021, ApJL, 913, L13, \dodoi{10.3847/2041-8213/abf11e}

\bibitem[{Ehlert {et~al.}(2023)Ehlert, Oikonomou, \& Unger}]{Ehlert_Curious_2023}
Ehlert, D., Oikonomou, F., \& Unger, M. 2023, PRD, 107, 103045, \dodoi{10.1103/PhysRevD.107.103045}

\bibitem[{Eichmann {et~al.}(2022)Eichmann, Kachelrie\ss{}, \& Oikonomou}]{Eichmann:2022ias}
Eichmann, B., Kachelrie\ss{}, M., \& Oikonomou, F. 2022, JCAP, 07, 006, \dodoi{10.1088/1475-7516/2022/07/006}

\bibitem[{Eichmann \& Winchen(2020)}]{eichmann_2020}
Eichmann, B., \& Winchen, T. 2020, JCAP, 2020, 047–047, \dodoi{10.1088/1475-7516/2020/04/047}

\bibitem[{Erdmann {et~al.}(2016)Erdmann, Müller, Urban, \& Wirtz}]{Erdmann_2016}
Erdmann, M., Müller, G., Urban, M., \& Wirtz, M. 2016, Astroparticle Physics, 85, 54–64, \dodoi{10.1016/j.astropartphys.2016.10.002}

\bibitem[{Farrar(2024)}]{farrar2024binary}
Farrar, G.~R. 2024.
\newblock \doarXiv{2405.12004}

\bibitem[{Farrar \& Sutherland(2019)}]{farrar_sutherland_deflections_2019}
Farrar, G.~R., \& Sutherland, M.~S. 2019, JCAP, 2019, 004, \dodoi{10.1088/1475-7516/2019/05/004}

\bibitem[{Globus {et~al.}(2023)Globus, Fedynitch, \& Blandford}]{Globus:2022qcr}
Globus, N., Fedynitch, A., \& Blandford, R.~D. 2023, Astrophys. J., 945, 12, \dodoi{10.3847/1538-4357/acaf5f}

\bibitem[{Globus {et~al.}(2019)Globus, Piran, Hoffman, Carlesi, \& Pomarède}]{Globus_2019}
Globus, N., Piran, T., Hoffman, Y., Carlesi, E., \& Pomarède, D. 2019, MNRAS, 484, 4167–4173, \dodoi{10.1093/mnras/stz164}

\bibitem[{{Golup, G. for the Pierre Auger Collaboration}(2023)}]{Golup_ADs_2023}
{Golup, G. for the Pierre Auger Collaboration}. 2023, in {PoS}({ICRC2023}), Vol. 444, 252, \dodoi{10.22323/1.444.0252}

\bibitem[{Gruppioni {et~al.}(2013)}]{Gruppioni_2013}
Gruppioni, C., {et~al.} 2013, MNRAS, 432, 23, \dodoi{10.1093/mnras/stt308}

\bibitem[{Harari {et~al.}(2000)Harari, Mollerach, \& Roulet}]{Harari_2000}
Harari, D., Mollerach, S., \& Roulet, E. 2000, JHEP, 2000, 035–035, \dodoi{10.1088/1126-6708/2000/02/035}

\bibitem[{Harari {et~al.}(2002)Harari, Mollerach, Roulet, \& Sánchez}]{Harari_2002}
Harari, D., Mollerach, S., Roulet, E., \& Sánchez, F. 2002, JHEP, 2002, 045–045, \dodoi{10.1088/1126-6708/2002/03/045}

\bibitem[{Ho(2008)}]{Ho_2008}
Ho, L.~C. 2008, Annual Review of Astronomy and Astrophysics, 46, 475, \dodoi{10.1146/annurev.astro.45.051806.110546}

\bibitem[{Hoffman {et~al.}(2018)Hoffman, Carlesi, Pomar{\`{e}}de, Tully, Courtois, Gottlöber, Libeskind, Sorce, \& Yepes}]{Hoffman_2018}
Hoffman, Y., Carlesi, E., Pomar{\`{e}}de, D., {et~al.} 2018, Nature Astronomy, 2, 680, \dodoi{10.1038/s41550-018-0502-4}

\bibitem[{Jaffe(2019)}]{Jaffe:2019iuk}
Jaffe, T.~R. 2019, Galaxies, 7, 52, \dodoi{10.3390/galaxies7020052}

\bibitem[{{Jansson} \& {Farrar}(2012{\natexlab{a}})}]{jansson_galactic_2012}
{Jansson}, R., \& {Farrar}, G.~R. 2012{\natexlab{a}}, ApJL, 761, L11, \dodoi{10.1088/2041-8205/761/1/L11}

\bibitem[{{Jansson} \& {Farrar}(2012{\natexlab{b}})}]{jansson_new_2012}
---. 2012{\natexlab{b}}, ApJ, 757, 14, \dodoi{10.1088/0004-637X/757/1/14}

\bibitem[{Korochkin {et~al.}(2024)Korochkin, Semikoz, \& Tinyakov}]{Korochkin:2024yit}
Korochkin, A., Semikoz, D., \& Tinyakov, P. 2024.
\newblock \doarXiv{2407.02148}

\bibitem[{Matthews {et~al.}(2018)Matthews, Bell, Blundell, \& Araudo}]{Matthews_2018}
Matthews, J.~H., Bell, A.~R., Blundell, K.~M., \& Araudo, A.~T. 2018, MNRAS: Letters, 479, L76–L80, \dodoi{10.1093/mnrasl/sly099}

\bibitem[{Murase \& Fukugita(2019)}]{Murase_2019}
Murase, K., \& Fukugita, M. 2019, PRD, 99.
\newblock \url{https://doi.org/10.1103%2Fphysrevd.99.063012}

\bibitem[{{The Pierre Auger Collaboration}(2014)}]{PierreAuger:2014gko}
{The Pierre Auger Collaboration}. 2014, Phys. Rev. D, 90, 122006, \dodoi{10.1103/PhysRevD.90.122006}

\bibitem[{{The Pierre Auger Collaboration}(2015)}]{auger_2015}
---. 2015, NIM A, 798, 172, \dodoi{10.1016/j.nima.2015.06.058}

\bibitem[{{The Pierre Auger Collaboration}(2017)}]{Auger_dipole_2017_Science}
---. 2017, Science, 357, 1266, \dodoi{10.1126/science.aan4338}

\bibitem[{{The Pierre Auger Collaboration}(2018)}]{Auger_dipole_2018}
---. 2018, ApJ, 868, 4, \dodoi{10.3847/1538-4357/aae689}

\bibitem[{{The Pierre Auger Collaboration}(2020)}]{Auger_spectrum_2020}
---. 2020, PRD, 102, 062005, \dodoi{10.1103/PhysRevD.102.062005}

\bibitem[{{The Pierre Auger Collaboration}(2022)}]{Auger_ADs_2022}
---. 2022, ApJ, 935, 170, \dodoi{10.3847/1538-4357/ac7d4e}

\bibitem[{{The Pierre Auger Collaboration}(2023)}]{Auger_CF_2023}
---. 2023, JCAP, 2023, 024, \dodoi{10.1088/1475-7516/2023/05/024}

\bibitem[{{The Pierre Auger Collaboration}(2024)}]{Auger_CFAD_2023}
---. 2024, JCAP, 2024, 022, \dodoi{10.1088/1475-7516/2024/01/022}

\bibitem[{{The Planck Collaboration}(2016)}]{Planck_2016}
{The Planck Collaboration}. 2016, A$\&$A, 596, A103, \dodoi{10.1051/0004-6361/201528033}

\bibitem[{Unger \& Farrar(2024)}]{UF23}
Unger, M., \& Farrar, G.~R. 2024, Astrophys. J., 970, 95, \dodoi{10.3847/1538-4357/ad4a54}

\bibitem[{{van Velzen} \& {Farrar}(2014)}]{vVfTDErate14}
{van Velzen}, S., \& {Farrar}, G.~R. 2014, \apj, 792, 53, \dodoi{10.1088/0004-637X/792/1/53}

\bibitem[{{Wanderman} \& {Piran}(2010)}]{grbRatePiran10}
{Wanderman}, D., \& {Piran}, T. 2010, MNRAS, 406, 1944, \dodoi{10.1111/j.1365-2966.2010.16787.x}

\end{thebibliography}
\bibliographystyle{aasjournal}



\end{document}